\documentclass[useAMS,usenatbib]{mn2e}

\usepackage{epsfig}

\def\pdeg{\ifmmode $\setbox0=\hbox{$^{\circ}$}\rlap{\hskip.11\wd0 .}$^{\circ}
          \else \setbox0=\hbox{$^{\circ}$}\rlap{\hskip.11\wd0 .}$^{\circ}$\fi}
\def\arcs{\ifmmode {^{\scriptscriptstyle\prime\prime}}
          \else $^{\scriptscriptstyle\prime\prime}$\fi}
\def\arcm{\ifmmode {^{\scriptscriptstyle\prime}}
          \else $^{\scriptscriptstyle\prime}$\fi}
\def\lesssim{\la}
\def\gtrsim{\ga}

\title[]{Spitzer IRS mapping of the central kpc of Centaurus A}

\author[]{
Alice C. Quillen$^{1}$,
Joss Bland-Hawthorn$^{2,3}$,
Joel Green$^{1}$,
\newauthor
J. D. Smith$^{4}$, 
D. Amelia Prasad$^{1}$,
Almudena Alonso-Herrero$^{5}$,
\newauthor
Mairi H. Brookes$^{6}$,
Kieran Cleary$^{6}$, \&
Charles R. Lawrence$^{6}$ \\
$^1${Department of Physics and Astronomy, University of Rochester, Rochester, NY 14627} \\
$^2${Anglo-Australian Observatory, P.O. Box 296, Epping NSW, Australia} \\
$^3${Institute of Astronomy, School of Physics, University of Sydney, NSW 2006, Australia} \\
$^4${Steward Observatory, University of Arizona, 933 North Cherry Avenue, Tucson, AZ 85721} \\
$^5${Departamento de Astrofisica Molecular e Infrarroja, Instituto de Estructura de la Materia, CSIC, E-28006 Madrid, Spain} \\
$^6${Jet Propulsion Laboratory, 4800 Oak Grove Drive, Pasadena, CA 91109} 
\\
{aquillen@pas.rochester.edu},
{jbh@aao.gov.au},
{joel@pas.rochester.edu},
{jdsmith@as.arizona.edu},
{aalonso@damir.iem.csic.es}
}

\begin{document}

\label{firstpage}

\maketitle

\begin{abstract}
We report on the results of spectroscopic mapping observations carried out
in the nuclear region of Centaurus A (NGC5128) over the 5.2 - 15 and 
20-36$\mu$m spectral regions using the Infrared Spectrograph 
on the {\it Spitzer Space Telescope}.  We have detected and mapped
S(0), S(2), S(3), and S(5) pure rotational transition lines 
of molecular hydrogen and emissions in the fine-structure transitions of 
[SiII], [SIII], [FeII], [FeIII], [ArII], [SIV], [NeII], [NeV] and [OIV].
The 500 pc bipolar dust shell discovered by Quillen et al.(2006)
is even more clearly seen in the 11.3$\mu$m dust emission feature
than previous broad band imaging.
The pure rotational lines of molecular hydrogen other than the
S(0) line are detected above the dusty disk and 
associated with the oval dust shell.
The molecular hydrogen transitions indicate the presence of warm gas
at temperatures 250--720K. The column density of the warm molecular hydrogen 
in the shell is $N(H_2) \sim 10^{20}$cm$^{-2}$ and 
similar to that estimated from the continuum dust shell surface brightness.
The ratio of the dust emission features at
7.7$\mu$m and  11.3$\mu$m  and the ratio of the
[NeII](12.8$\mu$m) and 11.3$\mu$m dust emission feature 
are lower in the 500 pc dust shell than in the star forming disk.  
The clearer shell morphology at 11.3$\mu$m,
warm molecular hydrogen emission in the shell,
and variation in line ratios in the shell compared to
those in the disk, confirm spectroscopically that
this shell is a separate coherent entity and is unlikely to
be a chance superposition of dust filaments.
The physical conditions in the shell are most similar to Galactic supernova
remnants where blast waves encounter molecular clouds.
The lines requiring the highest level of ionization,
[NeV]($24.318\mu$m) and [OIV]($25.890\mu$m),
are detected 20--25\arcsec north-east and 
south-west of the nucleus and at position angles near the radio jet axis.
Fine structure line ratios and limits from this region suggest that the medium 
is low density and illuminated by a 
hard radiation field at low ionization parameter. 
These higher S molecular hydrogen pure rotational transitions are 
also particularly bright in the same
region as the [OIV] and [NeV] emission.
This suggests that the gas associated with the dust shell has been 
excited near the jet axis and is part of an ionization cone.
\end{abstract}


\section{Introduction}
The nearest of all the giant radio galaxies, Centaurus~A
(NGC~5128) provides a unique opportunity to observe
the dynamics and morphology of an active galaxy
in detail across the electromagnetic spectrum.
For a recent review on this remarkable object see \citet{israel}.
In its central regions, NGC~5128 exhibits
a well recognized and optically-dark band of absorption
across its nucleus.
Images from the {\it Spitzer Space Telescope} (SST) with 
the Infrared Array Camera (IRAC) and Multiple Imaging Photometer for Spitzer
(MIPS) cameras
in the mid-infrared reveal a 3\arcmin long parallelogram shape
\citep{quillen_irwarp}
that has been modeled as a series of folds
in a dusty warped thin disk (e.g.,
\citealt{bland86,bland87,nicholson92,sparke96,quillen_irwarp}).
The dusty disk is also the site of ongoing star formation
at a rate of about $1M_\odot$~yr$^{-1}$,
(based on the infrared luminosity estimated by \citealt{eckart90}).
Centaurus~A hosts an active nucleus (e.g., \citealt{whysong04,mirabel99})
that has been
recently studied using infrared spectra from the SST.
\citep{weedman05,gorjian07}. Its nucleus exhibits a strong
silicate absorption feature and emission from [NeV].
Here we do not study the nucleus but focus on structure exterior to it.

The IRAC and MIPS images
of Centaurus~A have revealed another surprise, a bipolar
shell-like structure 500\,pc north and south of the nucleus 
\citep{quillen_supershell}.
This shell, seen for the first time with Spitzer imaging,  
could be the first extragalactic nuclear
shell discovered in the infrared.  

In this paper we present mapping of the central 500--600pc of
Centaurus A done with the Infrared Spectrograph (IRS) on board the SST.
The infrared spectral maps were obtained to test
the possibility that the apparent dust shell is a separate coherent
structure and not a chance superposition of dust filaments. 
They were also obtained to search for possible interactions
between the AGN and the interstellar medium
either by illumination from the AGN or
caused by jets or outflows.

Based on the discussion by \citet{israel}, we adopt a distance
to Cen A of 3.4~Mpc.  At this distance, 1~kpc 
corresponds to 1\arcmin on the sky.
All positions reported in this manuscript are given with respect to Epoch 2000.


\section{IRS observations}

Observations were obtained in spectral mapping mode in 
the Short-Low (SL;5--14$\mu$m) resolution and Long-High (LH; 19-38$\mu$m)  
resolution modules of the Infrared Spectrograph on board the SST.  
The spectral resolution of SL is $R \sim 50-100$ whereas that of
LH is $R\sim 600$. In spectral
mapping mode the Spitzer space craft points 
on a raster of discrete steps, settling
at each position before the integrations begin.
Each map was accompanied by a 4 position offset observation 
taken with the same slit pattern and exposure sequence. 
The offset (rogue) positions were used for sky subtraction.
The rogue sky position is  
at RA=13:25:15.7 (Epoch 2000) DEC=-42:55:13, approximately 6\arcmin
north of the nucleus and sufficiently far away that
little emission from the galaxy was seen in the 
previously observed broad band IRAC and MIPS images.

\subsection{SL observations}

We mapped in the short low (SL) module of the Infrared Spectrograph using
slit overlaps of 25\arcsec$\times$1.85\arcsec using a 60s ramp.
One data set covers $5\times 28$ map positions 
(through the guaranteed time observer program; GTO),
and the other two cover $7\times 30$ map positions 
(through the guest observer program; GO).
Both maps were observed by the SST on Mar 13, 2007.
Spectral cube assembly, including 
background subtraction, bad pixel removal, spatial regridding
and calibration was done using the software package
CUbe Builder for IRS Spectra Maps,
(CUBISM)\footnote{http://ssc.spitzer.caltech.edu/archanaly/contributed/cubism/}
\citep{smith07b} producing a single spectral cube.  

\subsection{LH observations}

We mapped in the long high spectral resolution (LH) module of the Infrared
Spectrograph using
slit overlaps of 10\arcsec$\times$4.5\arcsec, ~ so that
central positions in the map were covered by four slit positions.
We used a 60s ramp, and each observation contained 2 cycles.
For central map positions, 
this corresponds to a 480s on source integration time.
Two data sets (guest observer; GO) 
were taken covering  6$\times$20 map positions.
The third (guaranteed time observer; GTO) 
data set covers $5\times 16$ map positions.
All three maps were observed by the SST on Aug 1, 2006.

Data reduction of the three LH spectral data sets involved 
bad pixel removal, spectral
extraction, spatial regridding, and
application of the slit loss correction.   For descriptions
of the software procedures used see Appendix A by \citet{neufeld06}.
The resulting three data cubes were then resampled onto the same
spatial grid and averaged into a single cube, 
resulting in 1440s on source
exposure times for central positions in the map.

\subsection{Continuum morphology in the spectral cubes}

We have extracted continuum images from both spectral
cubes to check that they are consistent with previous broad band
imaging and to bring attention to features noted in previous
studies  \citep{quillen_irwarp,quillen_supershell} 
and that we will discuss below.

\begin{figure}
\includegraphics[angle=0,width=3.5in]{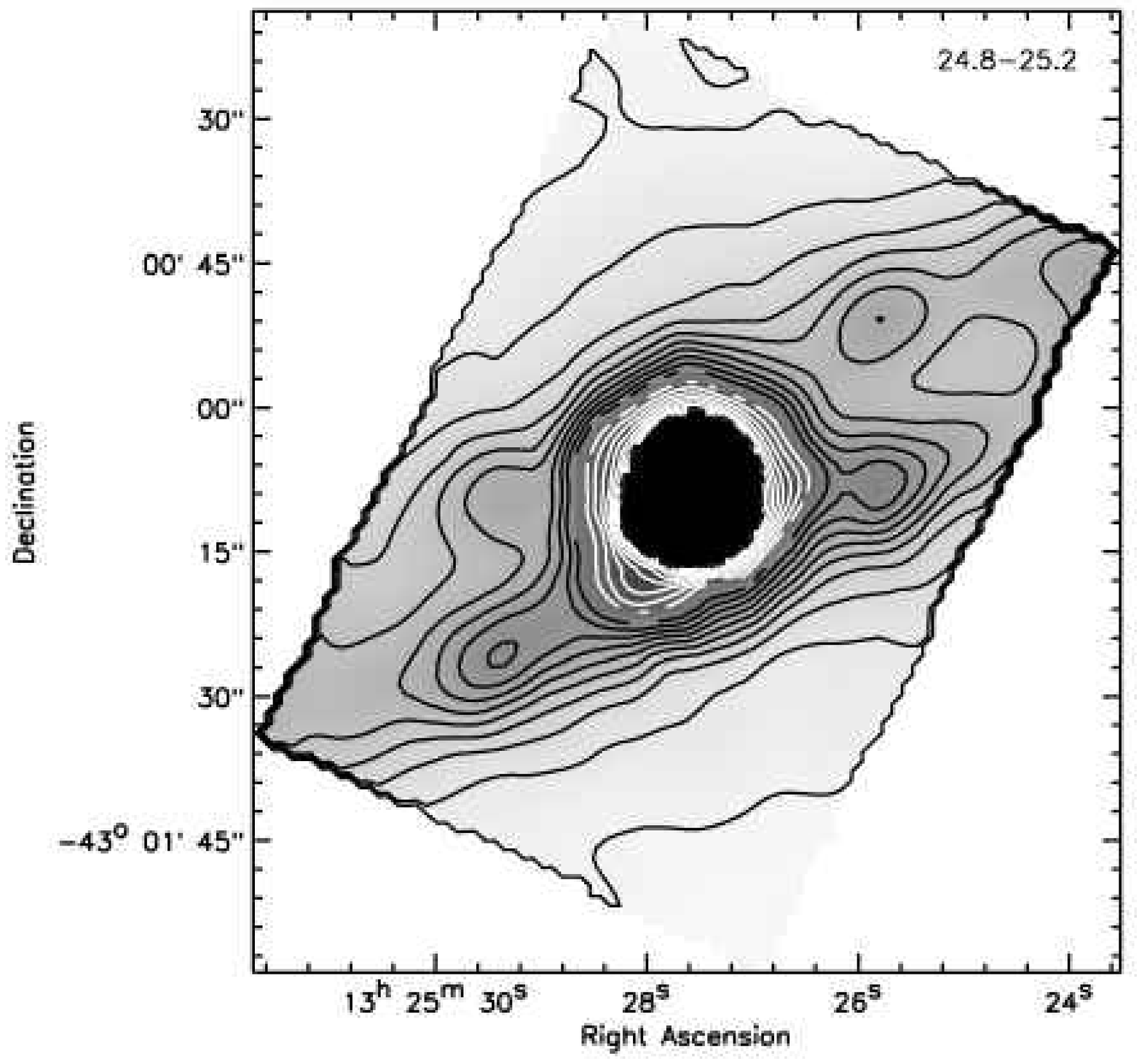}
\includegraphics[angle=0,width=3.5in]{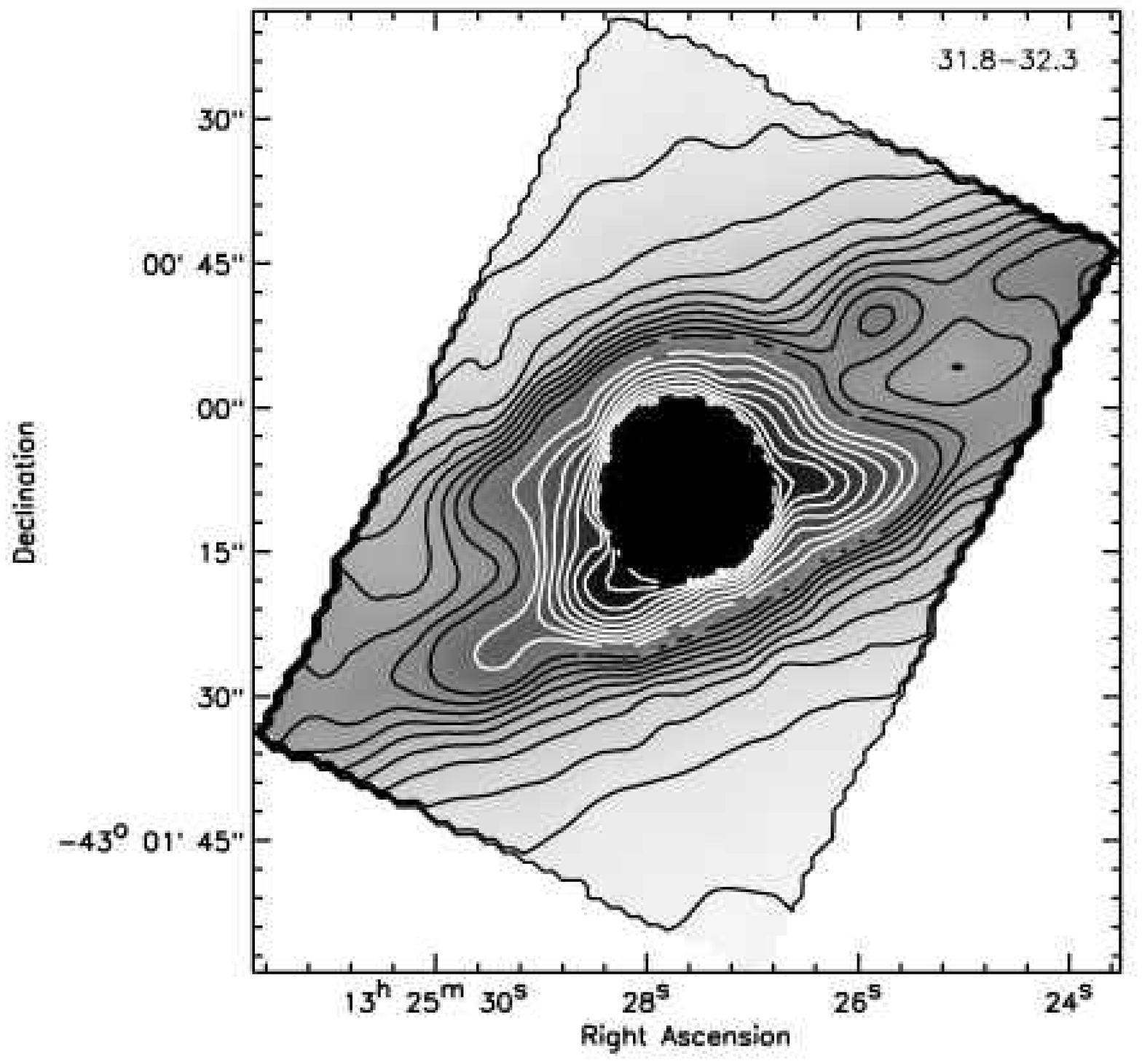}
\caption{
\label{fig:cont25}
Continuum emission from the LH spectral cube from $0.4\mu$m wide
bands.
Contours are evenly spaced with the lowest contour and spacing
at 0.01 MJy/SR.
a) Continuum centered at $25.0\mu$m.
b) Continuum centered at $32.0\mu$m.
}
\end{figure}

In Figure \ref{fig:cont25} we show the morphology of
the continuum emission in $0.4\mu$m wide bands
centered at 25$\mu$m and 32$\mu$m from the long-high spectral cube.  
We chose these regions so as to be free from artifacts caused
by order separation and emission lines.
The morphology at 25$\mu$m is consistent with the morphology
seen in the broad band MIPS $24\mu$m image previously
presented by \citep{quillen_irwarp}.
The broad band of emission approximately 30\arcsec wide,
and stretching approximately along the east-west direction
across the galaxy nucleus
we refer to as `the parallelogram' (see figure 3 by \citealt{quillen_irwarp}). 
The parallelogram shape is due to folds in a thin dusty warped 
disk. The warped disk is also rich in molecular gas \citep{quillenCO} and
so is the site of most active star formation present in the galaxy.

\begin{figure}
\includegraphics[angle=0,width=3.5in]{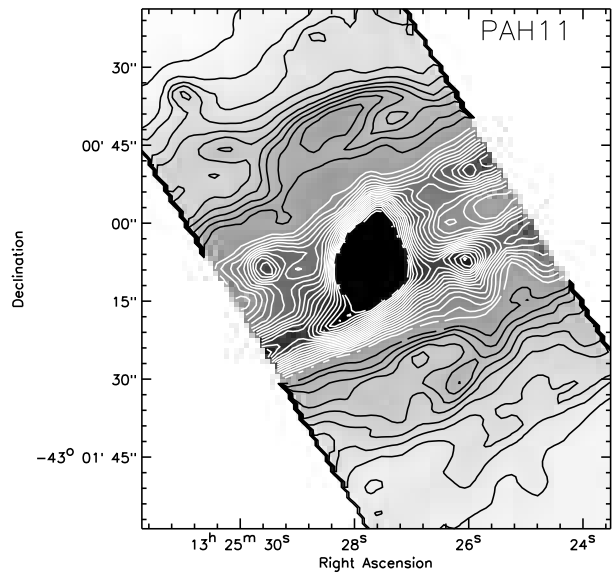}
\caption{
\label{fig:PAH11}
Dust emission feature at 11.3$\mu$m.
No continuum has been subtracted as the emission feature dominates
the spectrum by a factor of 3--8.
The minimum contour and spacing is approximately 
10 MJy/SR in the peak of the line.
The black contours show the oval dust shell previously
described by \citet{quillen_supershell}.
}
\end{figure}

In the SL spectral cube, broad and prominent dust emission features
are present. These dust emission features are generally attributed to 
polycyclic aromatic hydrocarbons (PAHs).
These dominate the broad
band emission in the IRAC broad band images previously studied 
\citep{quillen_supershell,quillen_irwarp} except possibly at the nucleus.
In Figure \ref{fig:PAH11} we show the morphology of the
peak of the $11.3\mu$m dust emission feature extracted from
the SL spectral cube.  This map also shows the parallelogram shape
or star forming disk seen in the IRAC band images. 
Above and below the parallelogram we can see the
shell-like bipolar structure described by \citet{quillen_supershell}
that was most prominent in the IRAC $8\mu$m image.
This shell has a major axis of $\sim 1'.1$, a minor axis
of $\sim 0'.7$ and position angle of 
approximately $10^\circ$ \citep{quillen_supershell}.
The IRAC 8.0$\mu$m band image is dominated by the PAH
dust emission feature at 7.7$\mu$m and so it is not surprising
that the 11.3$\mu$m image shows similar morphology. 
We show an image at 11.3$\mu$m here because it most clearly 
reveals the dust shell.  The morphology at 7.7$\mu$m and other PAH 
emission features and their ratios will be discussed in more detail below.
In the next section we will compare the morphology in line emission
maps to that seen in continuum (Figure \ref{fig:cont25})
and in the dust emission features.

In the $25.0\mu$m continuum image (Figure \ref{fig:cont25}a) 
three peaks in the emission are seen,
2 to the west of the nucleus, one to south-east.  A weaker but fourth is
located directly east of the nucleus.  These are labeled with arrows
on the 24$\mu$m  image in
Figure 2 by \citet{quillen_supershell}.
We bring these to the attention of the reader as we extract
spectra from these higher surface brightness regions to represent conditions 
typical of the star forming disk.  We will show that
spectra extracted from regions above or below the parallelogram region
differ from those extracted from the star forming disk or parallelogram
shaped region.

\section{Emission Line Maps}

Line emission images were created by summing emission
near the line peak and subtracting nearby continuum.
These images are shown in Figures \ref{fig:Ssi} -- \ref{fig:H2S0}
for lines present in the long wavelength spectral data cube 
and in Figures \ref{fig:H2S3} --  \ref{fig:NeII} for
lines or dust emission features present in the SL spectral data cube.
Line identifications and rest frame
wavelengths are summarized in Table \ref{tab:lines}.

\subsection{Morphology associated with the star forming warped disk}

\begin{figure}
\includegraphics[angle=0,width=3.5in]{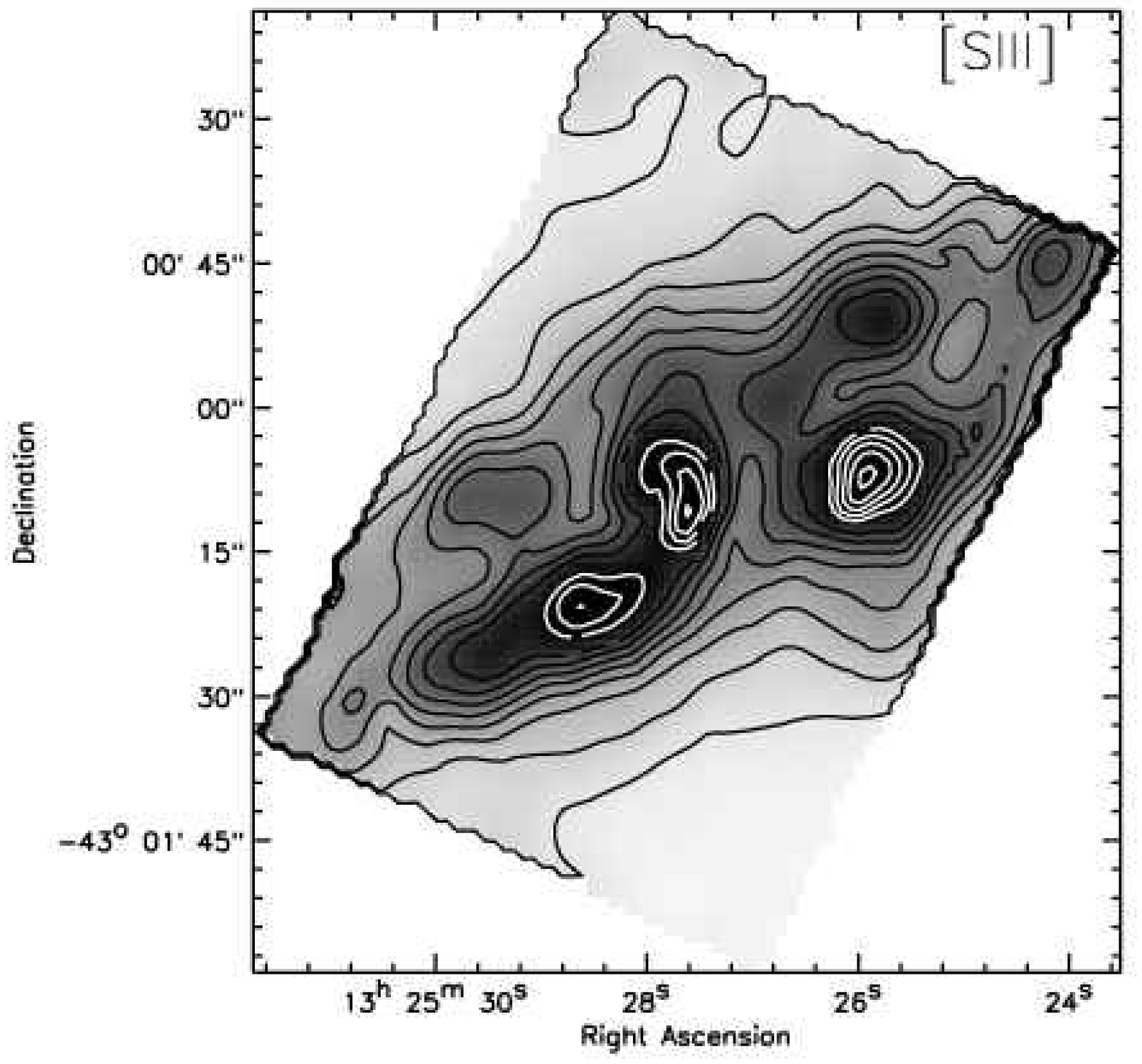}
\includegraphics[angle=0,width=3.5in]{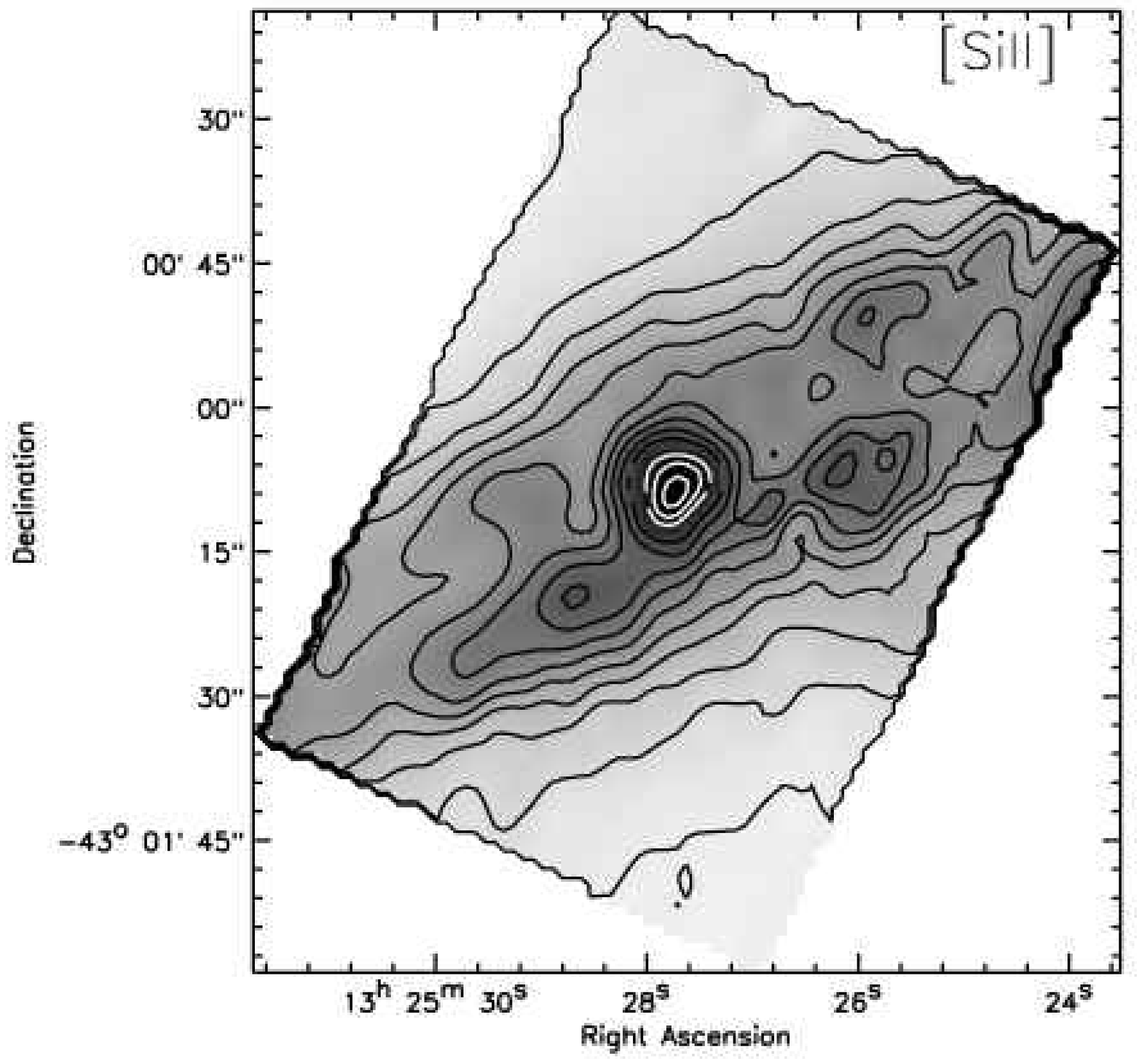}
\caption{
\label{fig:Ssi}
Continuum subtracted line emission images in the LH spectral cube,
showing emission in the folded star forming disk.
Contours are evenly spaced.
The lowest contours and spacing 
for the [SIII]($33.481\mu$m) and [SiII]($34.815$m) images
are 1.0  and 1.5  
$\times 10^{-8}$erg cm$^{-2}$ s$^{-1}$ SR$^{-1}$, respectively.
a) For [SIII]($33.481\mu$m).
b) For [SiII]($34.815\mu$m).
}
\end{figure}

\begin{figure}
\includegraphics[angle=0,width=3.5in]{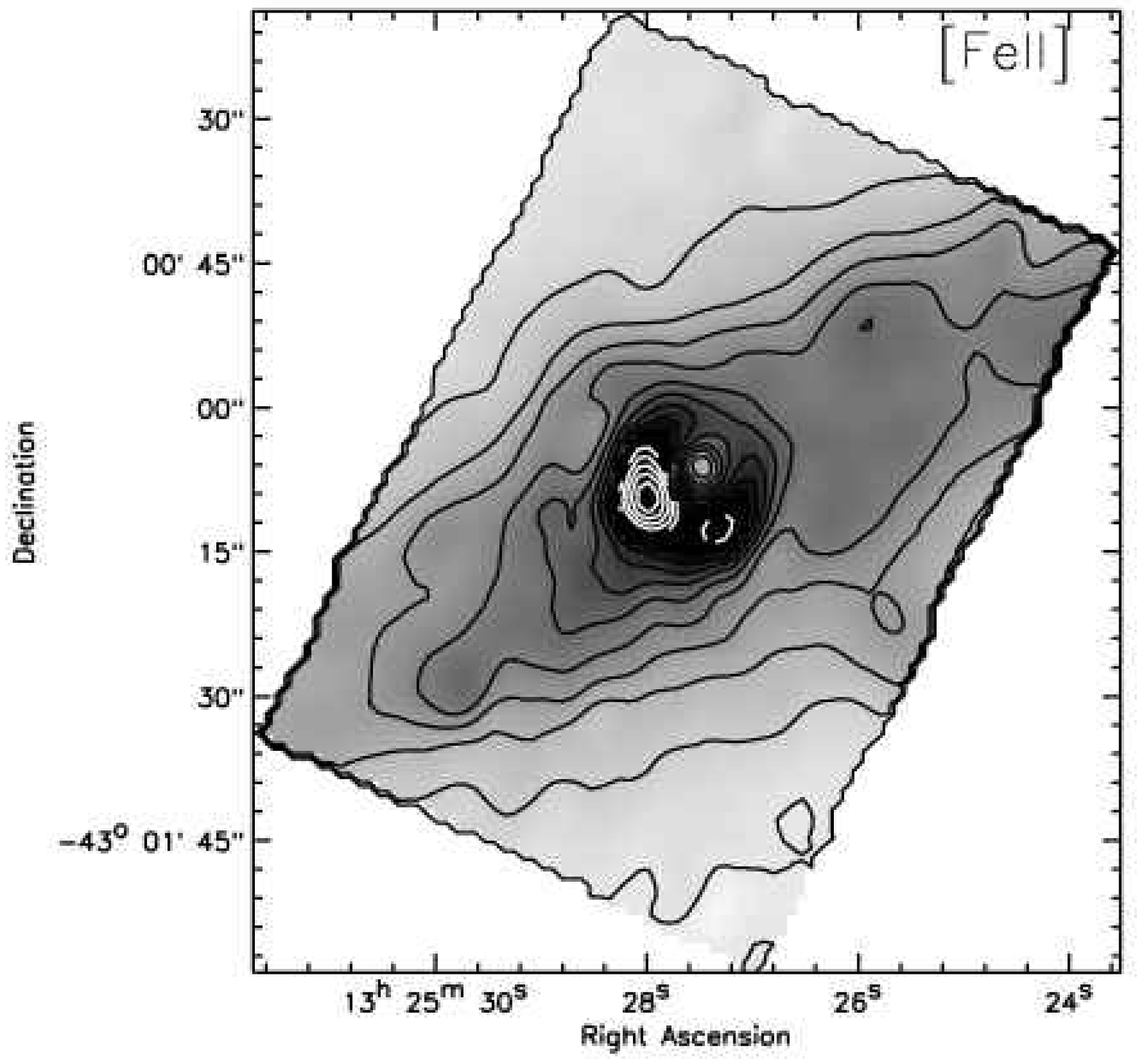}
\includegraphics[angle=0,width=3.5in]{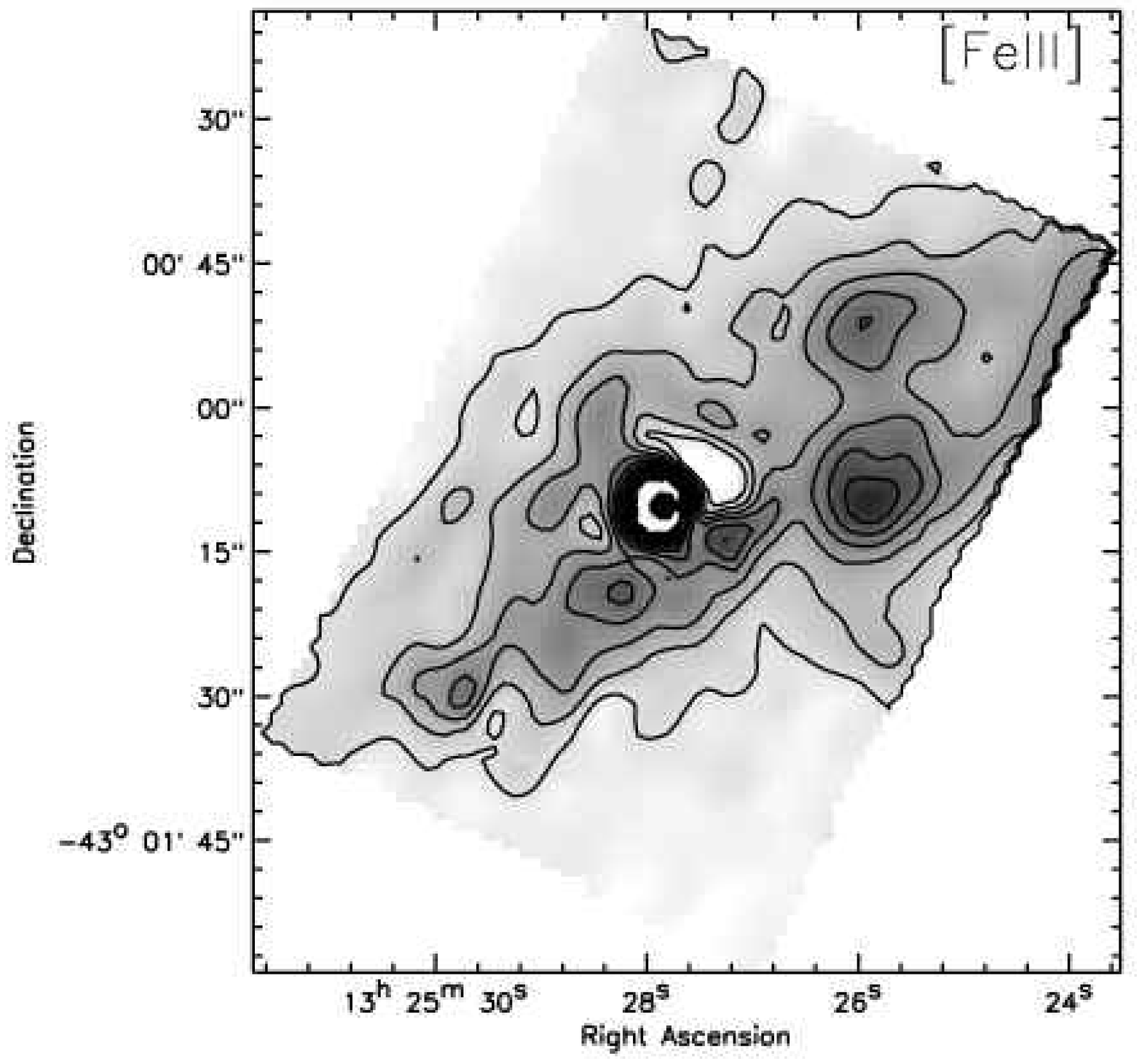}
\caption{
\label{fig:Fe}
Continuum subtracted line emission images in the LH spectral cube,
showing emission in the folded star forming disk.
Contours are evenly spaced.
The lowest contours and spacings  for
[FeII](25.988$\mu$m) and 
[FeIII](22.925$\mu$m) images with lowest contour are 
0.05  and 0.025
$\times 10^{-8}$erg cm$^{-2}$ s$^{-1}$ SR$^{-1}$, respectively.
a) For [FeII]($25.988\mu$m).
b) for [FeIII](22.925$\mu$m).
}
\end{figure}

\begin{figure}
\includegraphics[angle=0,width=3.5in]{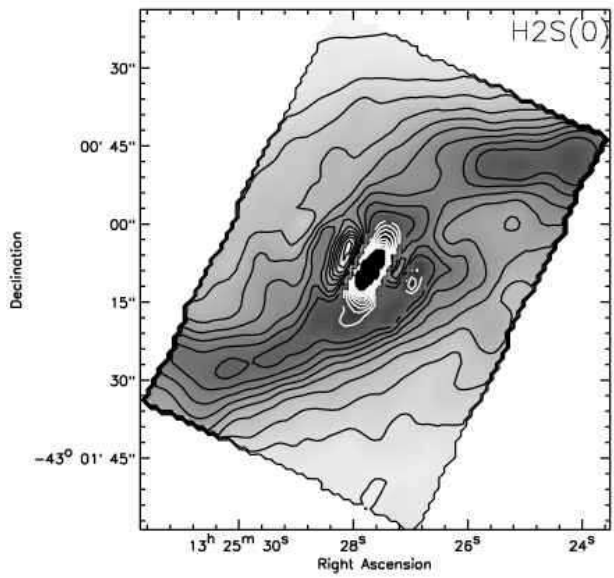}
\caption{
\label{fig:H2S0}
Continuum subtracted line emission images in the LH spectral cube,
showing emission in the folded star forming disk.
for the H$_2$S(0)($28.22\mu$m)  line
with lowest contour at and spacing at 0.01
$\times 10^{-8}$erg cm$^{-2}$ s$^{-1}$ SR$^{-1}$.
}
\end{figure}

The lines [SiII]($34.8\mu$m) and [SIII]($33.5\mu$m) 
(as displayed in Figure \ref{fig:Ssi})
are strong in the parallelogram or star forming disk,
as would be expected from the presence of HII regions in this disk.
Peaks in these maps are approximately coincident with peaks seen
in the nearby continuum at $32\mu$m  (Figure \ref{fig:cont25}b).
The [FeII]($26.0\mu$m) and [FeIII](22.93$\mu$m) lines are
also detected in the same region with [FeIII] weaker than [FeII]
(Figure \ref{fig:Fe}). 
Molecular hydrogen in the H$_2$S(0)$J=2$--0($28.2\mu$m) line 
is also detected associated with star formation in the warped
disk or parallelogram (Figure \ref{fig:H2S0}) though the morphology in this
line differs from that seen in the [SiII], [SIII], and [FeII] lines.
The northwestern and southeastern
ridges of the parallelogram are stronger than 
the southwestern and northern ridges.  The northwestern
and southeastern ridges also seem to be brighter in the
[FeII] map. 

The emission lines present in the long wavelength spectral 
cube are clearly associated with 
the disk, however there are differences between these maps
and the nearby continuum. 
There is more structure seen in the 
[SIII] map than present in nearby continuum at 32$\mu$m;
peaks in the disk are brighter than the surrounding emission
in the parallelogram (compare Figure \ref{fig:Ssi} to \ref{fig:cont25}b).  
\citet{quillen_supershell} suggested
that the 4 peaks in the broad band $24\mu$m images 
were due to a superposition of
the shell and the star forming disk.  However the emission lines
at long wavelengths are not 
detected in the vicinity of the shell-like structure but
do display the four emission peaks (particularly the [SIII] line image).
Therefore that explanation for the 4 peaks seen at 24$\mu$m is 
probably incorrect.
Those peaks were not predicted by the warped disk model presented by 
\citealt{quillen_irwarp}.  The peaks are symmetrical about
the nucleus suggesting that they are due to projection of a 
warped disk and that improvements in the warp model might succeed
in accounting for them.   If this were the case
then the size of the peaks should be similar in the continuum
and line emission.  Unfortunately the peaks are more prominent
in some of the lines (e.g., [SIII]) than the nearby continuum.
Spiral structure in the warped disk could
cause particular regions in the disk to exhibit 
higher levels of star formation and could maintain the reflective
symmetry.    
If this were the case then variations in the
observed morphology between lines and continuum would
require a model in which 
the strength of spiral structure depends on wavelength.

\subsection{[NeII] emission associated with the dust-shell}

The [SIII], [SiII] and [FeII] emission are primarily confined
to the warped disk (or parallelogram shape).
However the [NeII](12.8$\mu$m) image (see Figure \ref{fig:NeII}) 
does show faint levels of emission (at the
level of $\sim 10^{-7}$erg~cm$^{-2}$s$^{-1}$SR$^{-1}$)
associated with the dust shell. 
This line 
is blended with the nearby 12.7$\mu$m dust emission feature, nevertheless
the shape of the spectrum at different locations suggests that [NeII] dominates
the dust emission and is present
above and below the galaxy disk.  If so then the
[SIII], [SiII] and [FeII] emission associated with the dust shell is just too
faint to see in the long wavelength spectral cube. 
Deeper observations at longer wavelengths or higher spectral resolution short wavelength
infrared observations would be needed to confirm the presence
of ionized gas in the shell. A spectroscopic study at visible band wavelengths
would be quite sensitive as extinction is probably
not high in the shell region above and below the parallelogram. 

\subsection{Morphology associated with higher ionization lines, 
[OIV] and [NeV]}

\begin{figure}
\includegraphics[angle=0,width=3.5in]{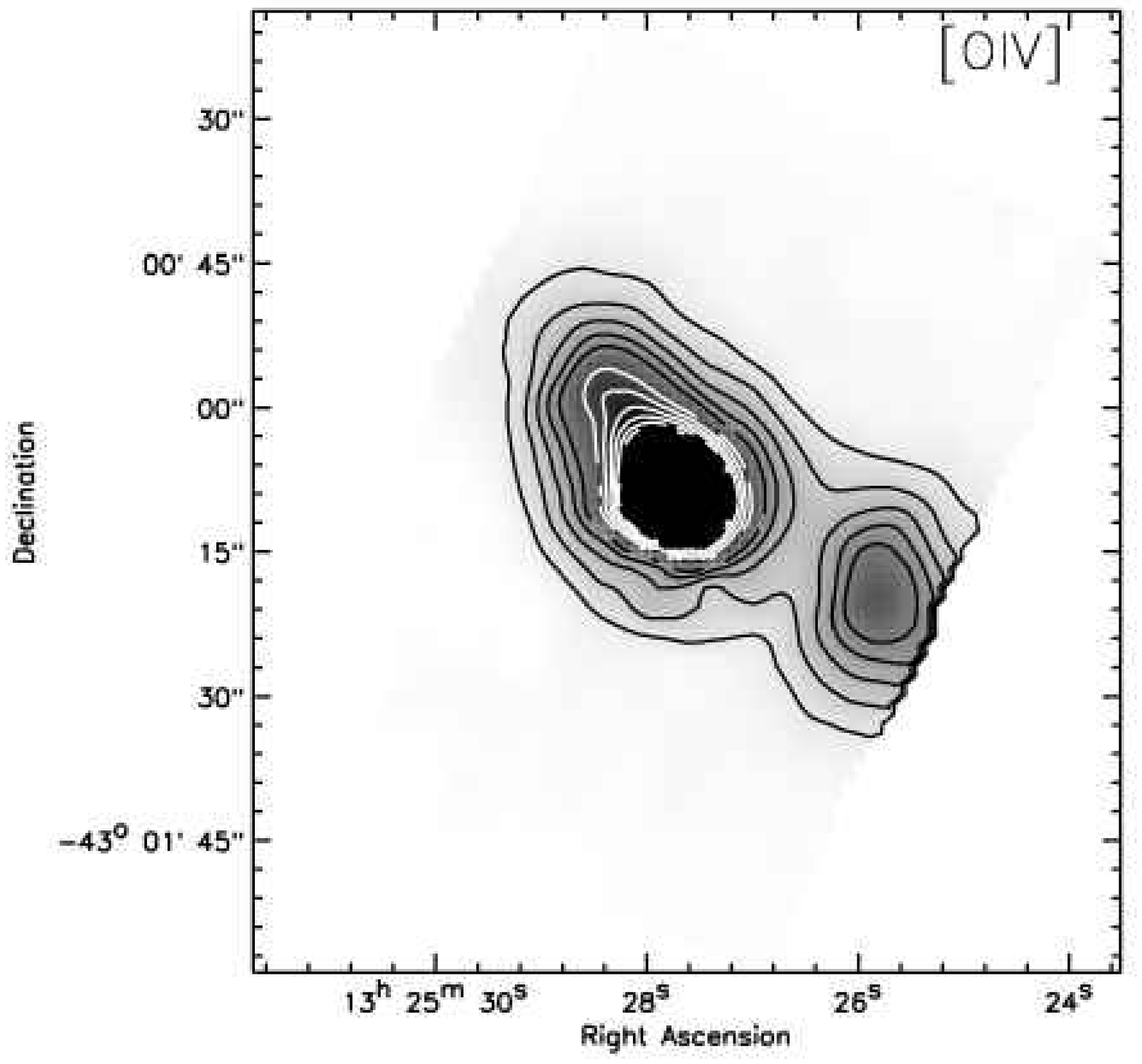}
\includegraphics[angle=0,width=3.5in]{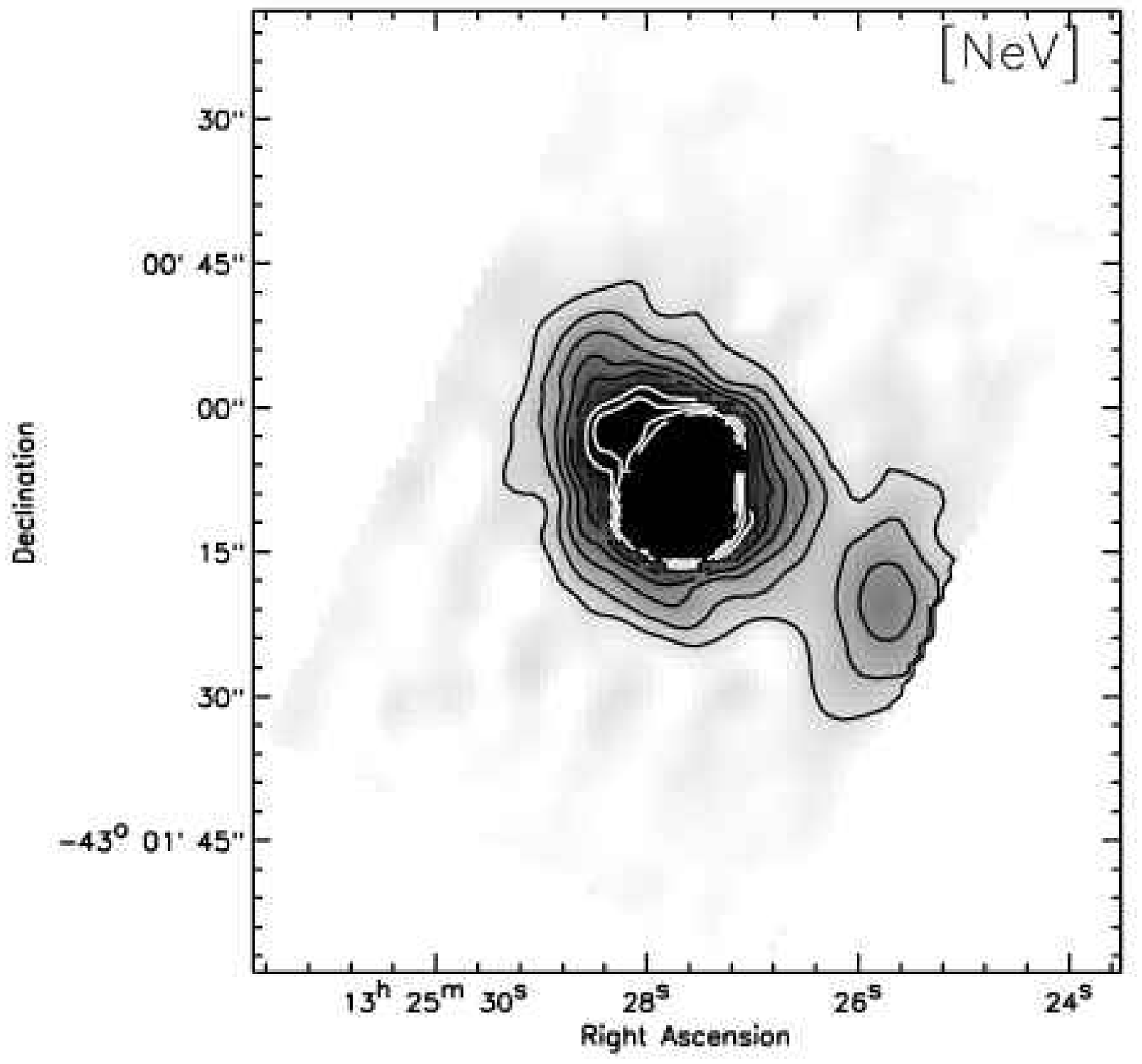}
\caption{
\label{fig:OIVNEV}
Continuum subtracted line emission images in the LH spectral cube,
showing emission near the jet axis.
Contours are evenly spaced.  The lowest contours and spacing for the 
[OIV]($25.890\mu$m) and [NeV]($24.318\mu$m) images
are 0.5 and 0.1  
$\times 10^{-8}$erg cm$^{-2}$ s$^{-1}$ SR$^{-1}$, respectively.
a) For [OIV]($25.890\mu$m).
b) For [NeV]($24.318\mu$m).
}
\end{figure}

\begin{figure}
\includegraphics[angle=0,width=3.5in]{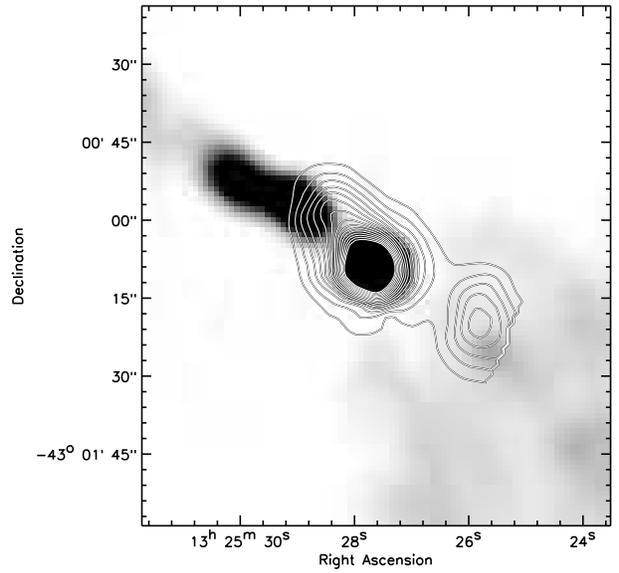}
\caption{
Radio emission at 5Ghz shown as grayscale
with [OIV]($25.9\mu$m) contours.
The radio map is a 6\arcsec resolution map of the inner radio lobes
by \citet{hardcastle06}.
The [OIV]$25.9\mu$ and [NeV]$24.3\mu$m line emission are oriented approximately
but not exactly along the jet axis.
\label{radioo4}
}
\end{figure}

Lines requiring higher ionization, 
[OIV]($25.9\mu$m) and [NeV]($24.3\mu$m), are also detected
in the LH spectral cube
and they exhibit different morphology than the
lower ionization lines.
Emission from
[OIV] and [NeV] is strong along position angles of $\sim 40^\circ$ and 
$-120^\circ$ from the nucleus and is resolved at locations
outside the nucleus;
see Figure \ref{fig:OIVNEV}.
These position angles are within 15$^\circ$
of the jet axis, though the jet is seen prominently only
on the north-east side in radio emission.  
The radio and X-ray jets in this region
have a position angle of $55^\circ$ \citep{burns83,kraft00}.
In figure \ref{radioo4} we show contours of radio continuum at 5 GHz
by \citet{hardcastle06}
overlayed on the [OIV]($25.9\mu$m) emission map (shown in grayscale).
We find that
the position angle of the [OIV] differs from the jet axis by $5-15^\circ$.
The different between the axis
of the [OIV] emission and radio jet axis 
is perhaps not surprising.  Observations
of Seyfert galaxies have found that the jet axis 
can differ from that of an ionization cone (NGC 1068, for example).

High angular resolution observations of the central few arcseconds
of Centaurus A
have revealed elongated or cone-shaped emission in Pa$\alpha$, Pa$\beta$ 
and [FeII](1.26$\mu$m) north of the nucleus 
that has been interpreted in terms of a rotating
disk, rather than in terms of an ionization cone \citep{schreier98,krajnovic06}.
The jet axis lies on the western side of the possible ionization cone
seen in [FeII](1.26).
The [OIV] and [NeV] extension to the north of
the galaxy nucleus are approximately oriented in the same
direction as the cone seen in [FeII](1.26$\mu$m)
shown by \citet{krajnovic06}.  As there is no cone feature seen
in the near-infrared [FeII] observations south of the nucleus (likely
because the band of absorption caused by the warped disk) it is more
difficult to compare the cone to the [OIV] emission south-west of
the nucleus. 
The [OIV] emission north and south
of the nucleus are not $180^\circ$ apart and the emission 
south-west of the nucleus would not overlay the [FeII](1.26) 
possibly ionization cone if rotated by $180^\circ$. 

It is interesting to reinspect the morphology of the 
lower ionization line maps at the position of the peak in [OIV]  south
west of the nucleus.   
An extension in the contours of the [SIII], [SiII], [FeII]
[FeIII] and H$_2$S(0) emission line brightness 
is seen at the same location as the peak
south-west of the nucleus in the [OIV] map.
The ratio of the surface brightness in the extension
compared to that in the disk is higher in the [FeIII] line
compared to the [FeII] line.   This is not surprising
as we know these extensions contain higher ionization species
such as [NeV].

The spectral resolution in the LH spectral cube  
(a spectral resolution of $R = {\Delta \lambda \over \lambda} \sim 600$ 
corresponds to a velocity resolution of $\Delta v = 500$km/s) 
is sufficiently high that we can just barely see the galactic disk's rotation
in the brighter lines (e.g., [SIII], [SiII] and H$_2$S(0)), with the
western and eastern sides of the disk redshifted or
blueshifted, respectively, compared to the system velocity.
We inspected the channels near the line centers and searched
for evidence of structure in the radial velocity distribution.
We found no clear evidence for a significant (greater than $\pm 200$km/s)
red or blue shift 
in the [OIV] emission extensions compared to the system velocity.
We also saw no clear evidence for an increase in line width.
The line width (FWHM) must be smaller
than the galactic rotation full width or $\lesssim 400$km/s. 
Near the [OIV] peak south-west of the nucleus
the extensions in the contours of the [SIII], [SiII], 
and H$_2$S(0) also appear to have velocities
within 200km/s of the system velocity. 
Unfortunately the spectral resolution is just not
high enough to be certain about 
these line widths or allow a search for structure in
the radial velocity distribution.

\subsection{Emission from molecular hydrogen}

\begin{figure}
\includegraphics[angle=0,width=3.5in]{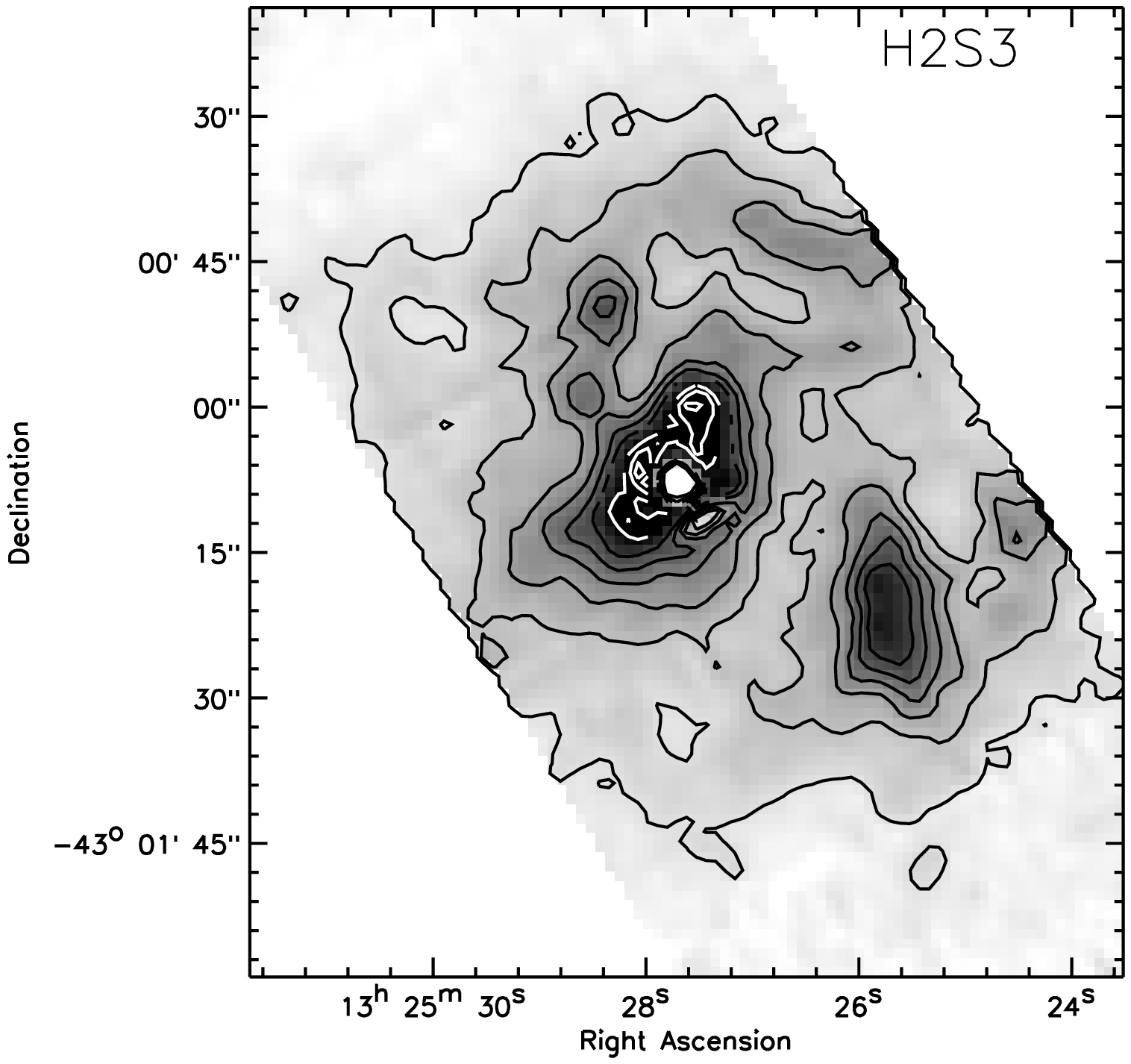}
\includegraphics[angle=0,width=3.5in]{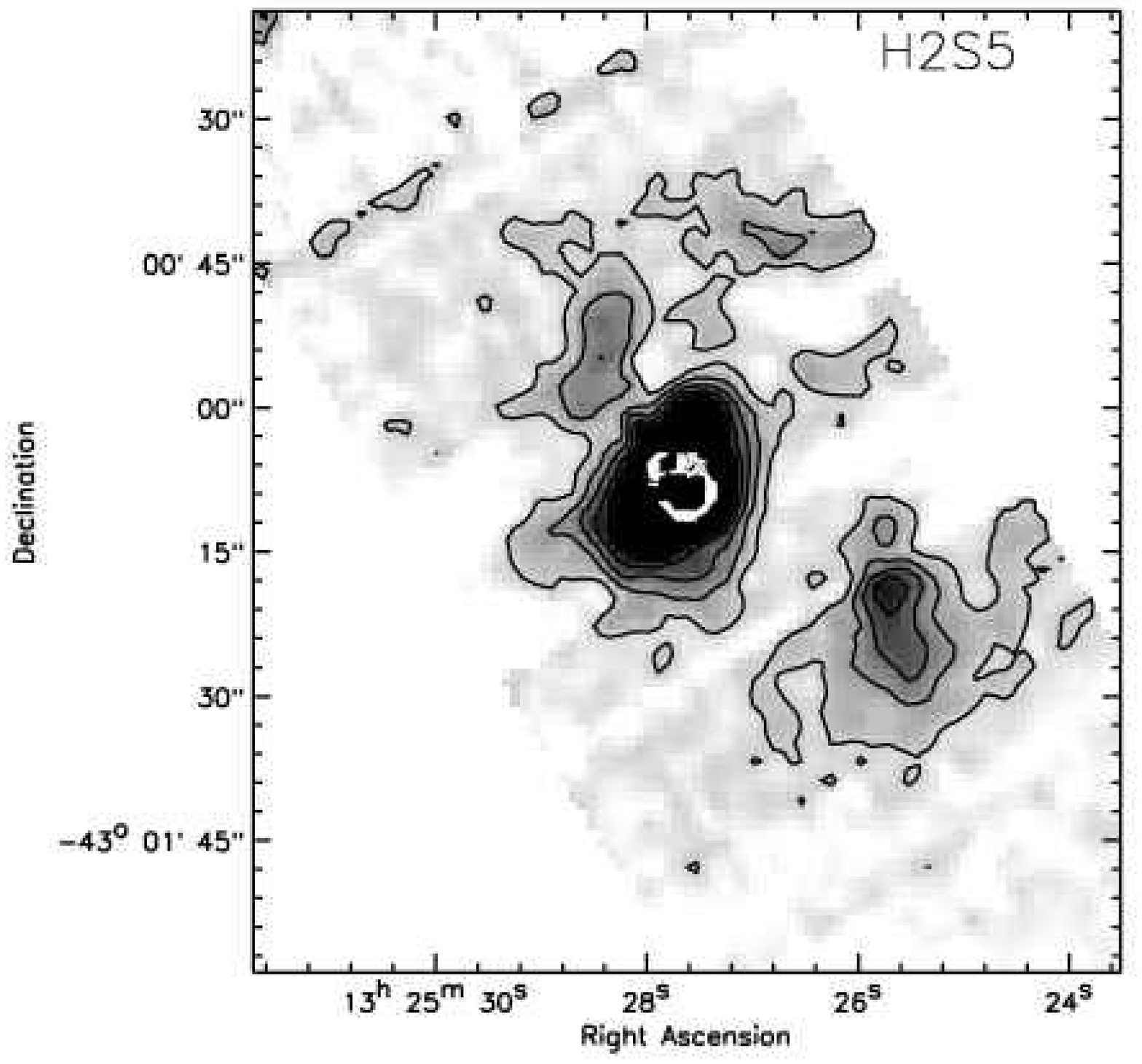}
\caption{
\label{fig:H2S3}
Continuum subtracted line emission images from the SL spectral cube, showing
the 
H$_2$S(3)(9.665$\mu$m) and H$_2$S(5)(6.909$\mu$m) lines with lowest contours at 
1 $\times 10^{-7}$erg cm$^{-2}$ s$^{-1}$ SR$^{-1}$.
The higher S rotational lines from H$_2$ exhibit different morphology than
the H$_2$S(0)28$\mu$m line that was seen primarily in the folded star forming disk.
Emission in the higher S lines is seen above the disk.
Contours are evenly spaced.  
Lowest contour and spacing is  $\times 10^{-7}$erg cm$^{-2}$ s$^{-1}$ SR$^{-1}$ 
for both lines.
For a) H$_2$S(3)(9.665$\mu$m).
For b) H$_2$S(5)(6.909$\mu$m).
}
\end{figure}

In the SL spectral cube
we detect four pure rotational molecular hydrogen lines, 
the S(7)$J=7$--5, the S(5)$J=5$--3, the S(3)$J=3$--1 
and the S(2)$J=2$--0 lines.   
We show the continuum subtracted line emission maps 
for the H$_2$S(5)(6.909$\mu$m)
and  H$_2$S(3)(9.665$\mu$m) lines in Figure \ref{fig:H2S3}.  
The S(7) is weak and the S(2) strongly
blended with dust emission features so we have only presented
line emission maps from the other two lines.
The S(4) line at 8.0251$\mu$m was not detected, probably due to the 
bright dust emission feature 
at 7.7$\mu$m at its expected wavelength
and because the even number transitions are singlets
rather than triplets and so have one third the
number of states.  The S(1) line at 17.03$\mu$m 
was not detected because it lies outside the spectral range covered
by the SL and LH modules.  To observe this line we would have needed
observations with the SH module of the Infrared Spectrograph.

\begin{figure}
\includegraphics[angle=0,width=3.5in]{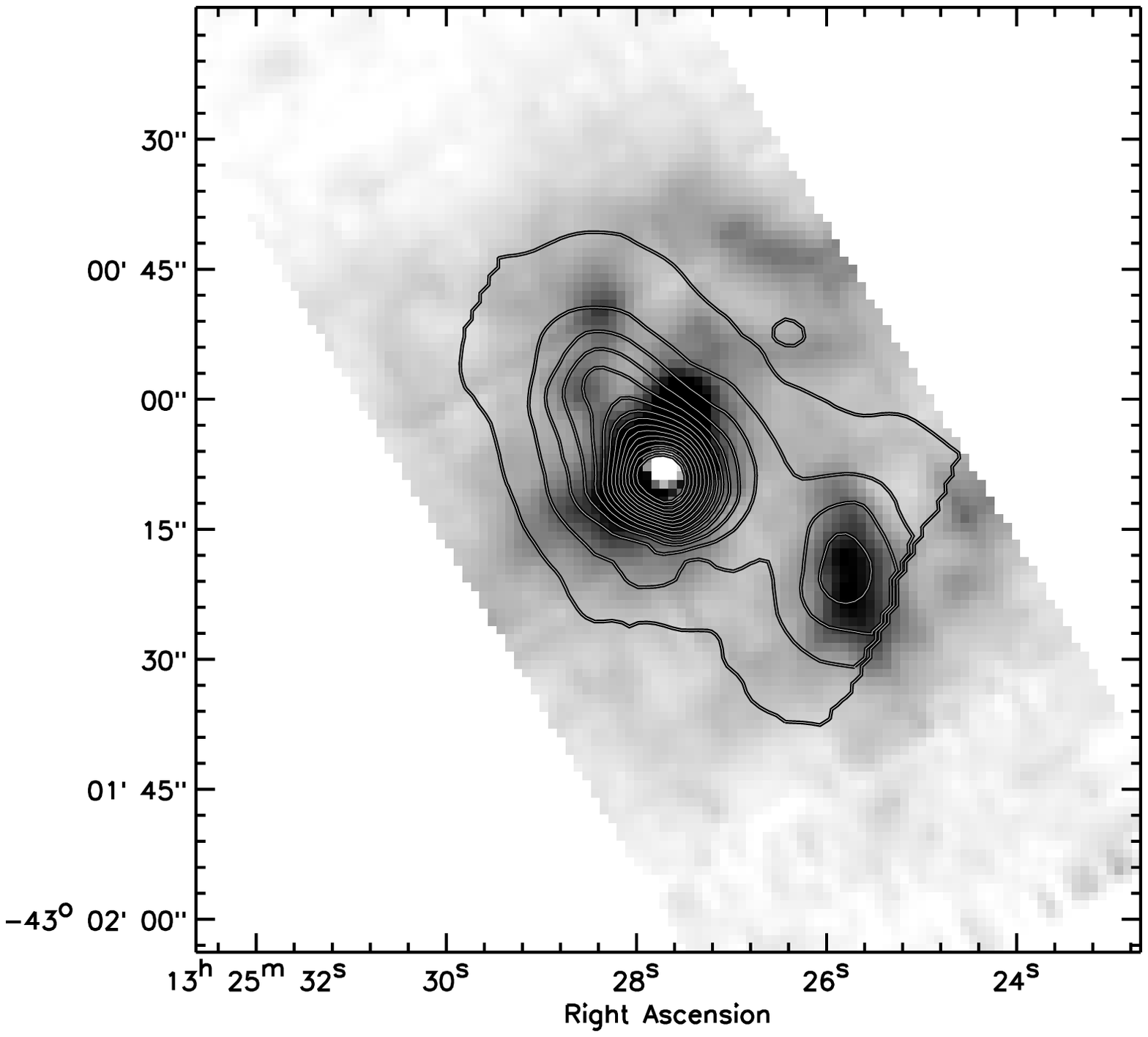}
\includegraphics[angle=0,width=3.5in]{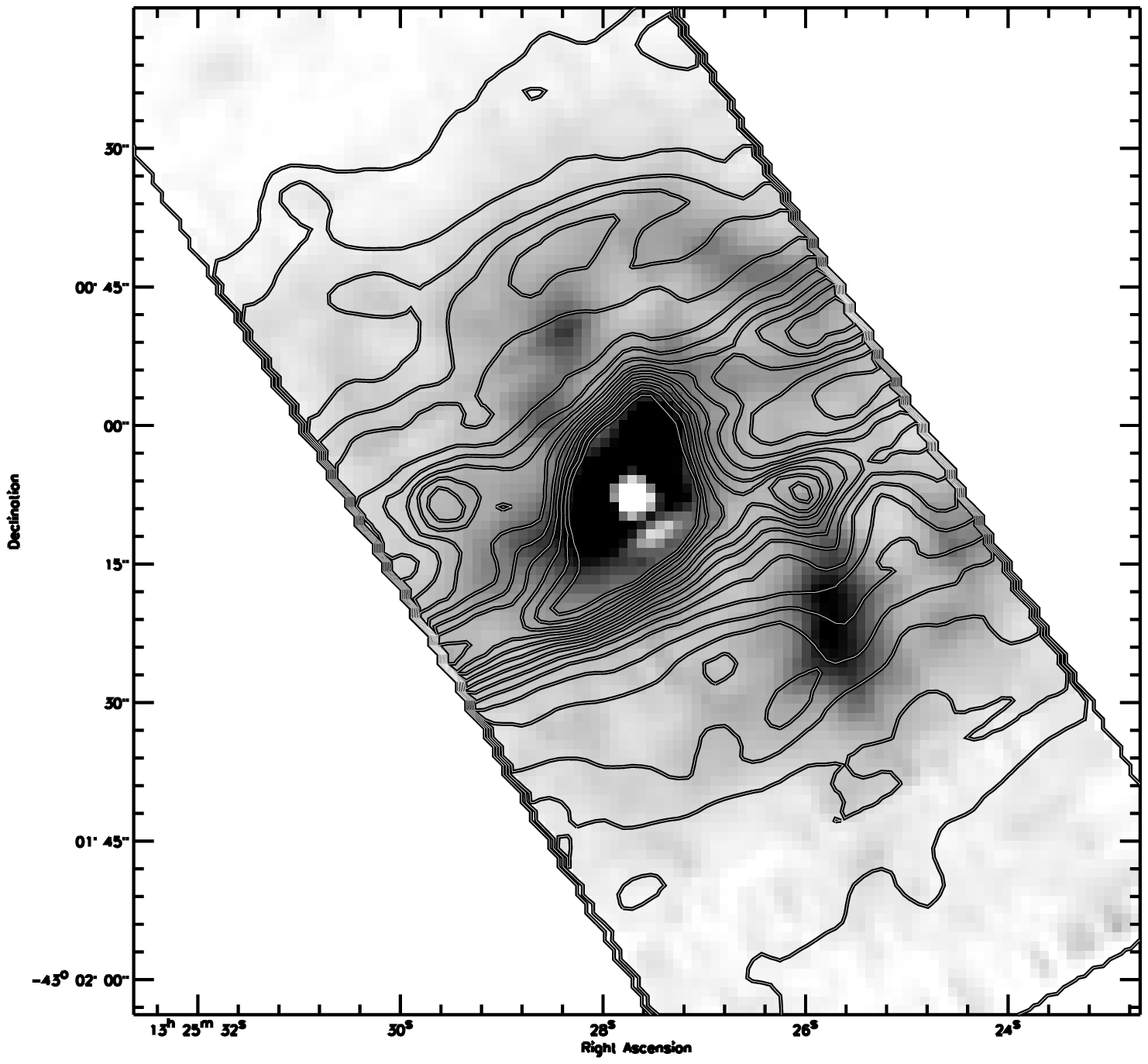}
\caption{
\label{fig:h2s3_ov}
a) Emission in H$_2$S(3)(9.665$\mu$m) shown as gray scale overlayed with
[OIV]($25.890\mu$) contours.
Contours are evenly spaced with lowest contours at 
are 0.5 $\times 10^{-8}$erg cm$^{-2}$ s$^{-1}$ SR$^{-1}$
and a spacing 4 times this.
The gray scale range for the H$_2$S(3) image is 0(white) to 7(black) 
$\times 10^{-7}$erg cm$^{-2}$ s$^{-1}$ SR$^{-1}$.
b) Same as a) except the H$_2$S(3) emission is overlayed with contours
of the 11.3$\mu$m  PAH dust emission feature (as shown in figure \ref{fig:PAH11}).
Contour spacing and lowest level is 20 MJy/SR.
}
\end{figure}

\begin{figure}
\includegraphics[angle=0,width=3.5in]{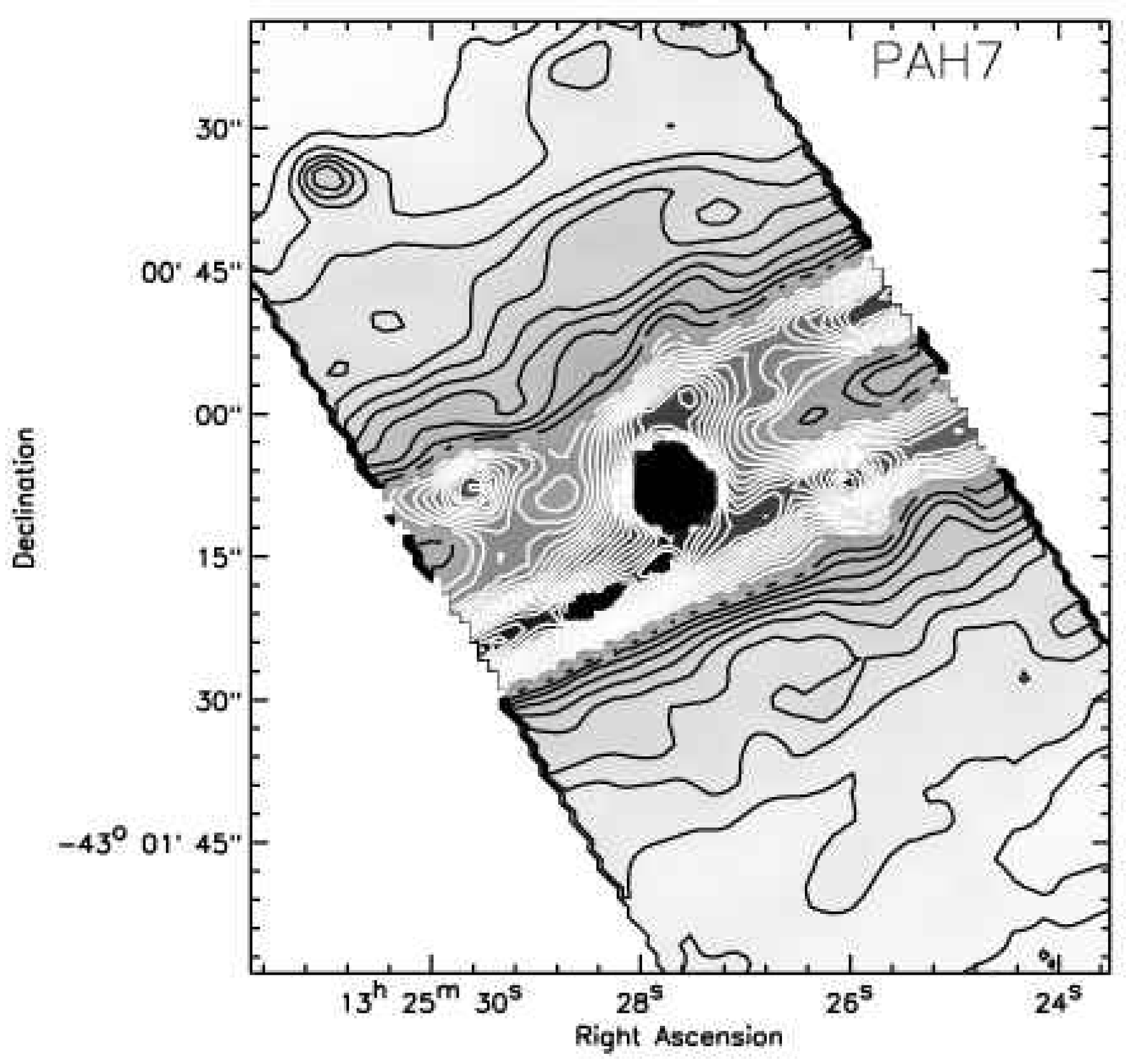}
\includegraphics[angle=0,width=3.5in]{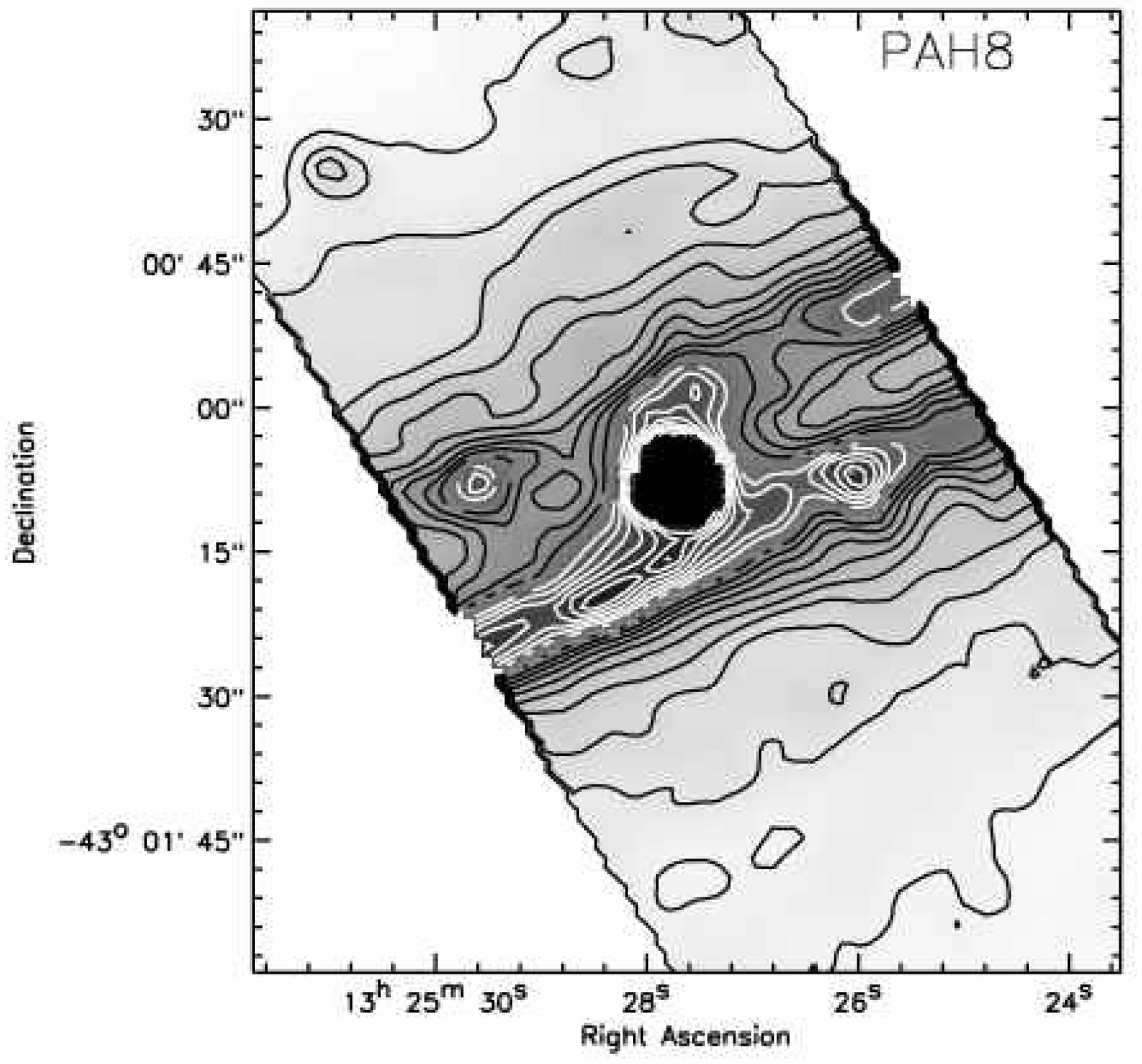}
\caption{
\label{fig:PAH}
Flux at 7.7 and 8.6$\mu$m showing dust emission features.
Contours are evenly spaced.  No continuum subtraction has been done.
The star forming disk is evident as the parallelogram shaped feature
corresponding to the folded disk.
The dust shell is seen above  and below the parallelogram feature.
The minimum contour and spacing is approximately 
10 MJy/SR in the peak of the line.
a) The 7.7$\mu$m dust emission feature.
b) The 8.6$\mu$m dust emission feature.
}
\end{figure}

\begin{figure}
\includegraphics[angle=0,width=3.5in]{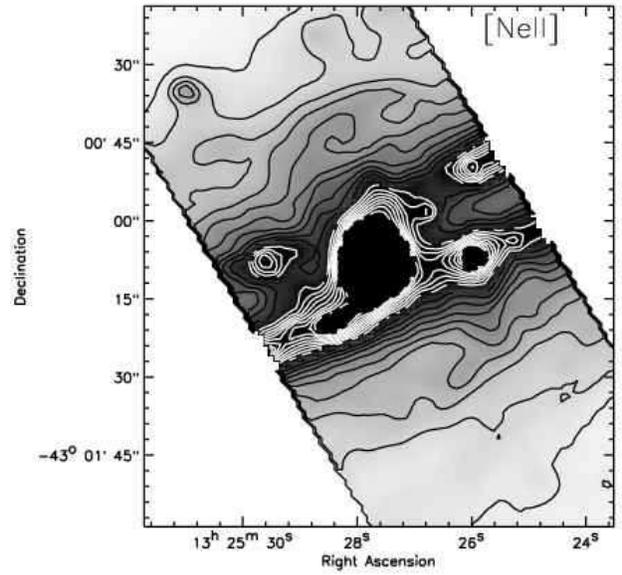}
\caption{
\label{fig:NeII}
Line emission in [NeII]($12.81\mu$m).
No continuum has been subtracted as
the [NeII] line dominates the continuum by a factor greater than 2 everywhere.
Contours are evenly spaced with lowest contour at 
4 $\times 10^{-7}$erg cm$^{-2}$ s$^{-1}$ SR$^{-1}$.
}
\end{figure}

The morphology of the higher S(3) and S(5) rotational molecular 
hydrogen lines differs
from that we see at longer wavelengths in the S(0) line at 28$\mu$m
(compare Figure \ref{fig:H2S0} to Figure \ref{fig:h2s3_ov}).  
The S(0) line is primarily associated with the star forming disk, however the 
S(3) and S(5) lines are seen above the disk.
An overlay of the H$_2$S(3)(9.665$\mu$m) line and [OIV](25.89$\mu$m) 
in Figure \ref{fig:h2s3_ov}a
shows that the peak in the [OIV] south-west of the nucleus coincides 
with the peak H$_2$S(3) surface brightness in the same region. 
An overlay shown in Figure \ref{fig:h2s3_ov}b 
of the S(3) line with the dust emission in the 11.3$\mu$ PAH emission
feature (previously shown in Figure \ref{fig:PAH11}) shows that
the warmer molecular hydrogen gas (traced by the higher S pure rotational
transitions) is near the dust shell.
The simplest explanation for the morphology 
is that warmer molecular hydrogen is associated with the
dust shell but that there is additional excitation of the
shell material near the jet axis.
Constraints on the column depth and temperature of
the molecular hydrogen are discussed below in section 5.

\subsection{Colors in the dust-shell compared to the warped disk}

\begin{figure}
\includegraphics[angle=0,width=3.5in]{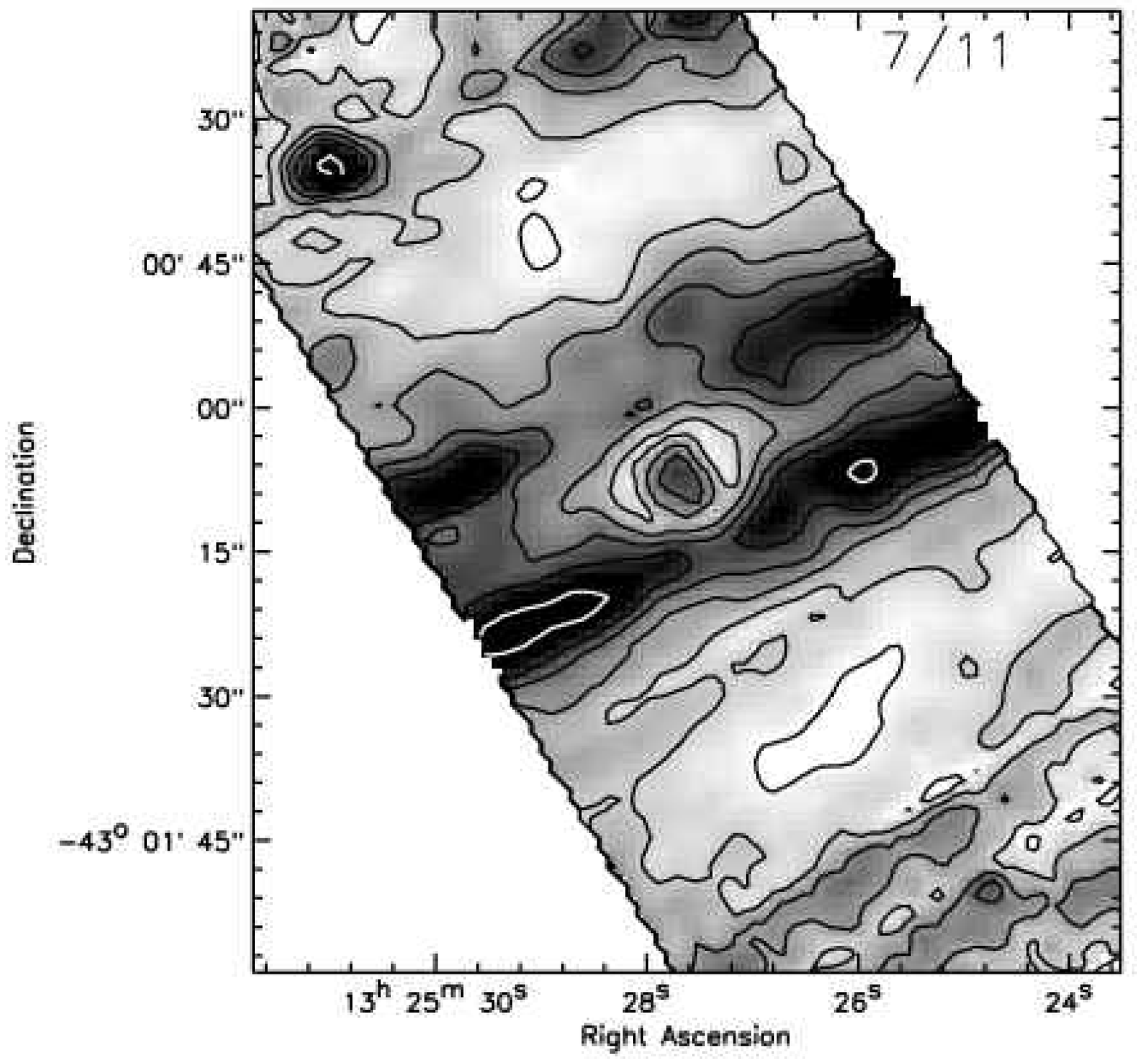}
\includegraphics[angle=0,width=3.5in]{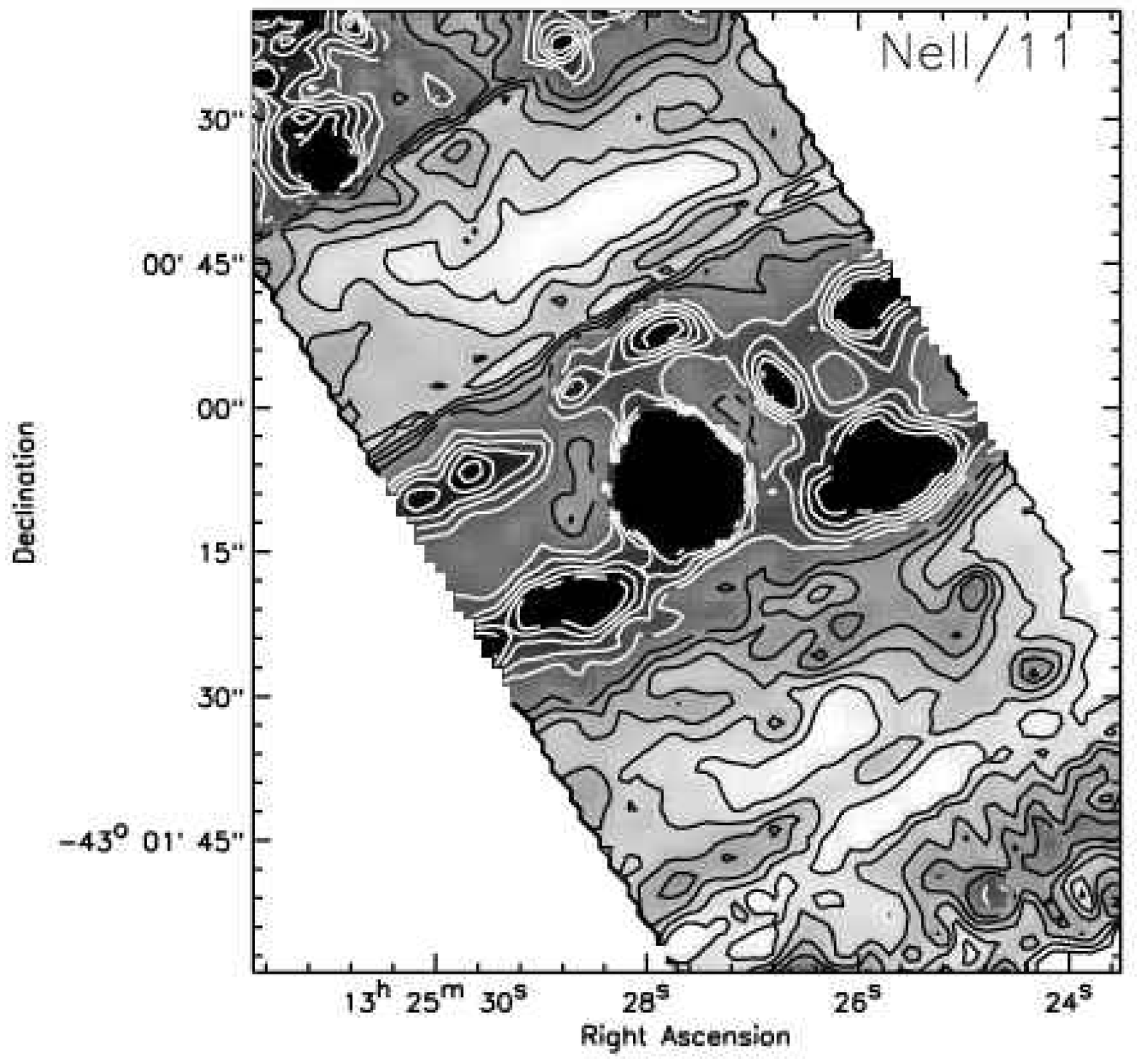}
\caption{
\label{fig:ratio}
a) The dust emission feature at 7.7$\mu$m divided by that
at 11.3$\mu$m.  
The lowest and highest contours
are shown at a ratios of 0.5 (in shell) and 1.0 (in parallelogram),
with spacing of 0.1.  
Black is 1.0, white is 0.5.
The 7.7 to 11.0$\mu$m dust feature ratio varies with the dust
shell having the lowest ratio.  
b) The [NeII](12.8$\mu$m)
line divided by the $11\mu$m dust emission feature.
The lowest and highest contours
are shown at ratios of 1.9 (in shell) and 3.0 (in parallelogram), 
with spacing of 0.1.
Black is 3.0, white is 1.9.
The strength of the NeII](12.8$\mu$m) compared to the 11.3$\mu$m 
dust feature also varies with the dust shell having the lowest ratios. 
}
\end{figure}

In Figure \ref{fig:PAH} and Figure \ref{fig:NeII} we show maps of continuum emission
at the peak of the 7.7, and 8.6$\mu$m dust emission features
and the [NeII]12.81$\mu$m line.  We did not subtract continuum from these maps as
the [NeII] line and dust emission feature are 3-8 times brighter than the continuum.  
As expected, the [NeII] line
and dust emission features primarily trace the star forming regions in the warped disk.
Fainter levels of emission arise from the dust shell above and
below the disk  and from the region that is bright in [OIV] emission 
($\sim$25\arcsec south
west of the nucleus).
In Figure \ref{fig:ratio} we show ratios of the 7.7$\mu$m and 
11.3$\mu$m surface brightness and the ratio of the   
[NeII]12.8 and 11.3$\mu$m surface brightness.  Both of these ratios
exhibit variations across the galaxy, with the lowest values 
in the dust shell and higher values present in the star forming disk.
Subtraction of the continuum made little difference in the morphology of
the ratio maps
because the lines and dust emission features are over twice as strong as the continuum.

\citet{smith07} found a correlation between the 7.7/11.3$\mu$m ratio of dust
emission complexes
and the [NeIII](15.6)/[NeII](12.8$\mu$m) ratio in star  forming galaxies.
They also noted that the galaxies with low 7.7/11.3 complex ratios 
were more likely to be classified as LINERS or Seyferts.
They suggested that the 7.7$\mu$m/11.3$\mu$m ratio is an indicator 
of the hardness of the radiation field.  
The [NeIII](15.6$\mu$m) line lies is outside the wavelength coverage of 
the SL module so we cannot use that ratio to probe the radiation
field above the disk.    At the nucleus in Cen A, this line was detected but
was not strong \citep{weedman05}.
We do have higher ionization tracers in the 
the [OIV](25.9$\mu$m) and [NeV](24.3$\mu$m) lines.  However, the morphology
in those lines differs from that we see in the 
the 7.7 to 11.3$\mu$m ratio map. 
The [OIV] line is bright near the jet axis whereas
whereas the 7.7/11.3 ratio is 
low in the dustshell (see Figure \ref{fig:ratio}).
The lack of change in the 7.7/11.3 ratio in the region
containing [OIV]25.9 and [NeV]24.3 emission near the jet axis makes it difficult
to interpret variations
in the 7.7/11.3 ratio purely in terms of hardness of radiation field.
Since the 7.7/11.3 ratio drops in the vicinity of the
dust shell and this is where we detect warm  molecular gas
we might instead consider models in which PAH emission is
affected by the presence of warm gas or the physical
process that is responsible for heating the warm gas. 
We could also consider models where this ratio
is affected by the ambient radiation field strength.

\section{Spectra}

We have extracted spectra at various positions in the star forming disk
from the LH spectral cube 
and these are shown in figure \ref{fig:LHspec}a.  
Spectra were extracted 
from 7\arcsec square regions
centered at 5 regions.  4 of these
correspond to peaks evident in Figure \ref{fig:Ssi} excluding
the nucleus.   From bottom to top the spectra 
shown in Figure \ref{fig:LHspec}a were extracted from regions 
centered at
RA=13:25:25.8, DEC=-43 00 52  (peak to the west and north of the nucleus),
RA=13:25:25.9, DEC=-43:01:09  (peak west of the nucleus),
RA=13:25:29.5, DEC=-43:01:27  (south-east of the nucleus),
RA=13 25 29.5, DEC=-43:01:10  (peak east of the nucleus), and 
RA=13:25:28.6, DEC=-43:01:23  (peak south-east of the nucleus).

We also extract
an LH spectrum along the jet axis at the peak
shown in Fig.\ref{fig:LHspec}b.
This spectrum was extracted over a 11\arcsec square region centered at
RA=13:25:25.7, DEC=-43:01:24, at the peak in the [OIV] emission 
south-west of the nucleus (see figure \ref{fig:OIVNEV}).
The lines we have detected from these
spectra are listed in Table \ref{tab:lines}.  
Line fluxes measured from this region are listed in Tables \ref{tab:emjet} and
\ref{tab:hjet}.

We see only minor differences in different
regions of the parallelogram (see Figure \ref{fig:LHspec}).
However, along the jet axis there are deviations in the spectra,
in the relative strengths of the lines requiring
higher levels of ionization, 
[OIV]($25.9\mu$m) and [NeV]($24.3\mu$m), compared
to those requiring lower levels (e.g., [SiII]($34.8\mu$m) ,[SIII]($33.5\mu$m)).
To produce the emitting ion from the preceding ionization state, 
[OIV] requires an ionization energy of 54.0ev where as
[NeV] requires 97.1ev.  
A harder UV radiation field  most naturally accounts
for these species.

\begin{figure}
\includegraphics[angle=0,width=3.5in]{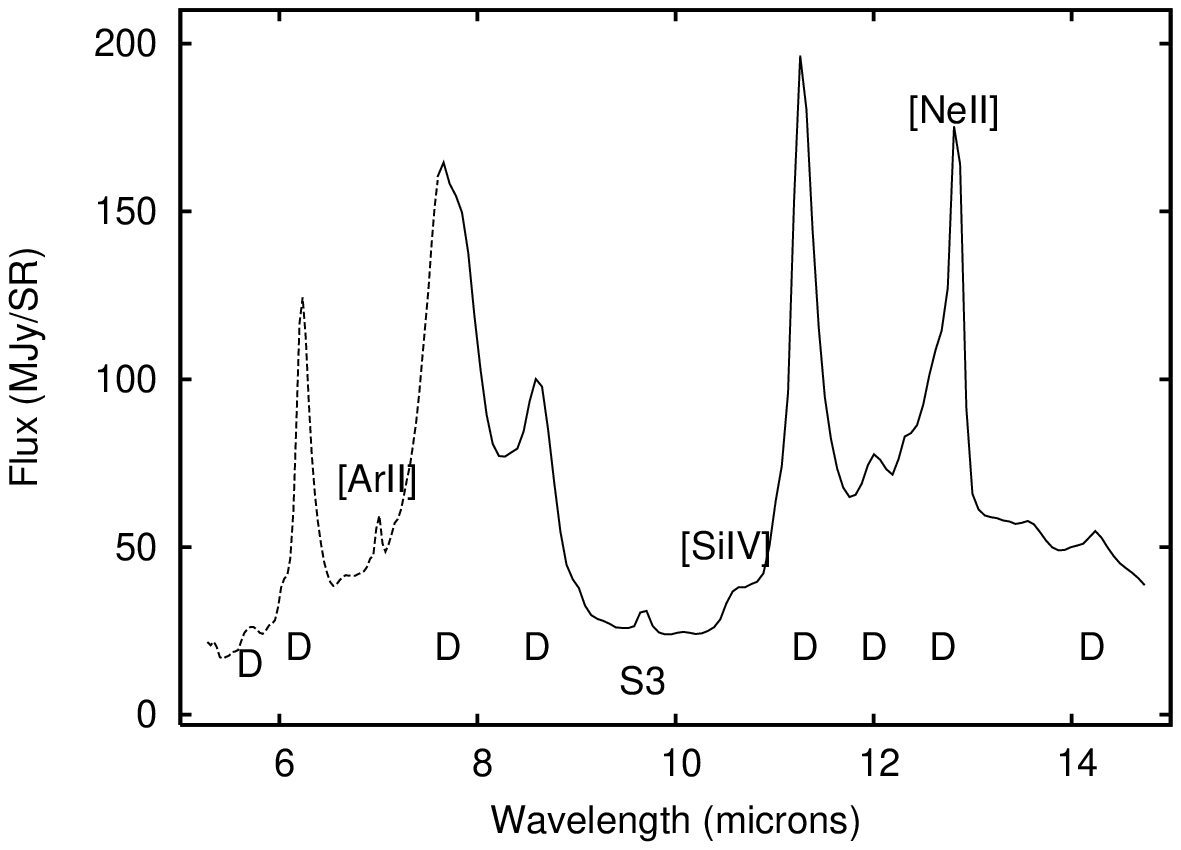}
\includegraphics[angle=0,width=3.5in]{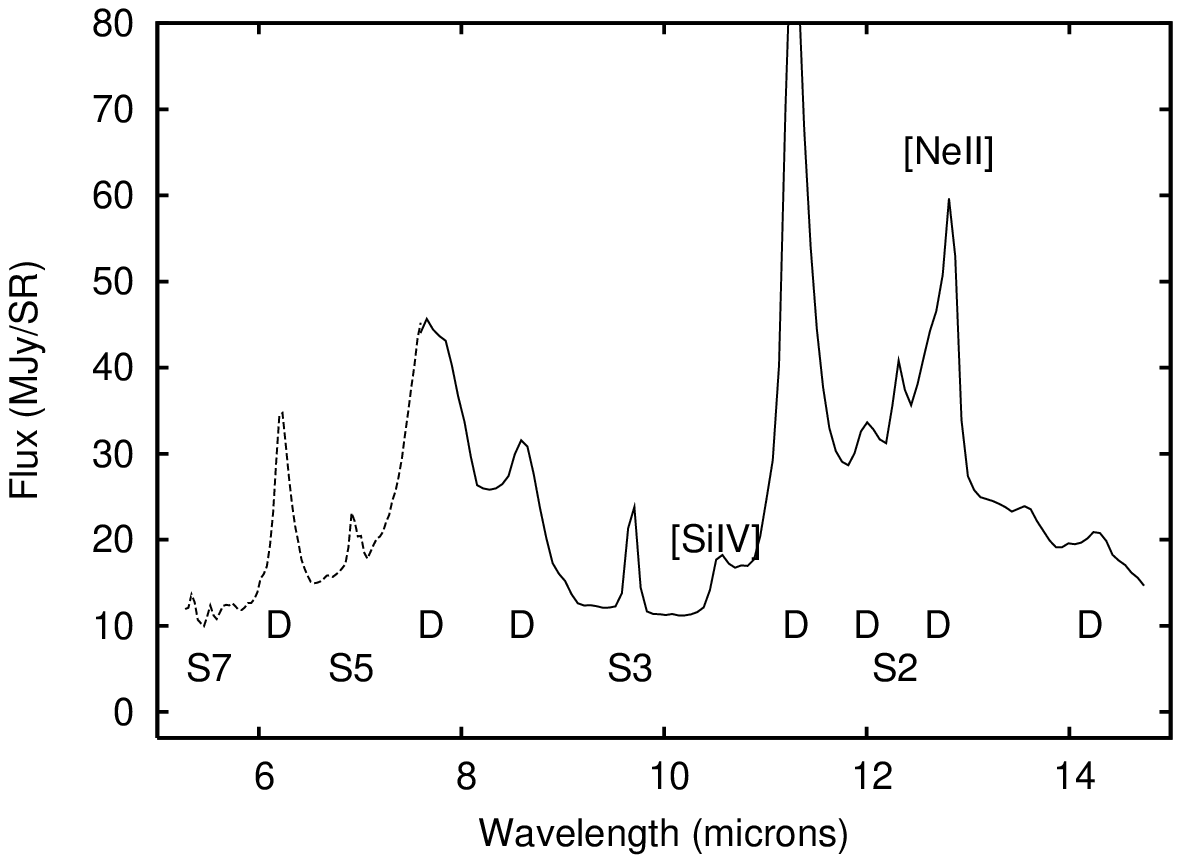}
\caption{
\label{fig:SLspec}
a) Spectrum from the star forming parallelogram.
D refers to a dust emission feature.
b) Spectrum from the jet region.
The pure rotational molecular hydrogen lines S(2)--S(7) are labeled as S2-S7.
Molecular hydrogen lines are more prominent in the jet region than
in the star forming parallelogram.
In the star forming disk the [ArII](6.985$\mu$m) line is brighter 
than the nearby 
pure rotational molecular hydrogen S(2) line at $6.909\mu$m  whereas
the opposite is true in the jet region.
There is a change in the ratio of the dust emission features
in these two spectra and in the ratio of the [NeII]/11.3 dust emission feature.
}
\end{figure}

\begin{figure}
\includegraphics[angle=0,width=3.5in]{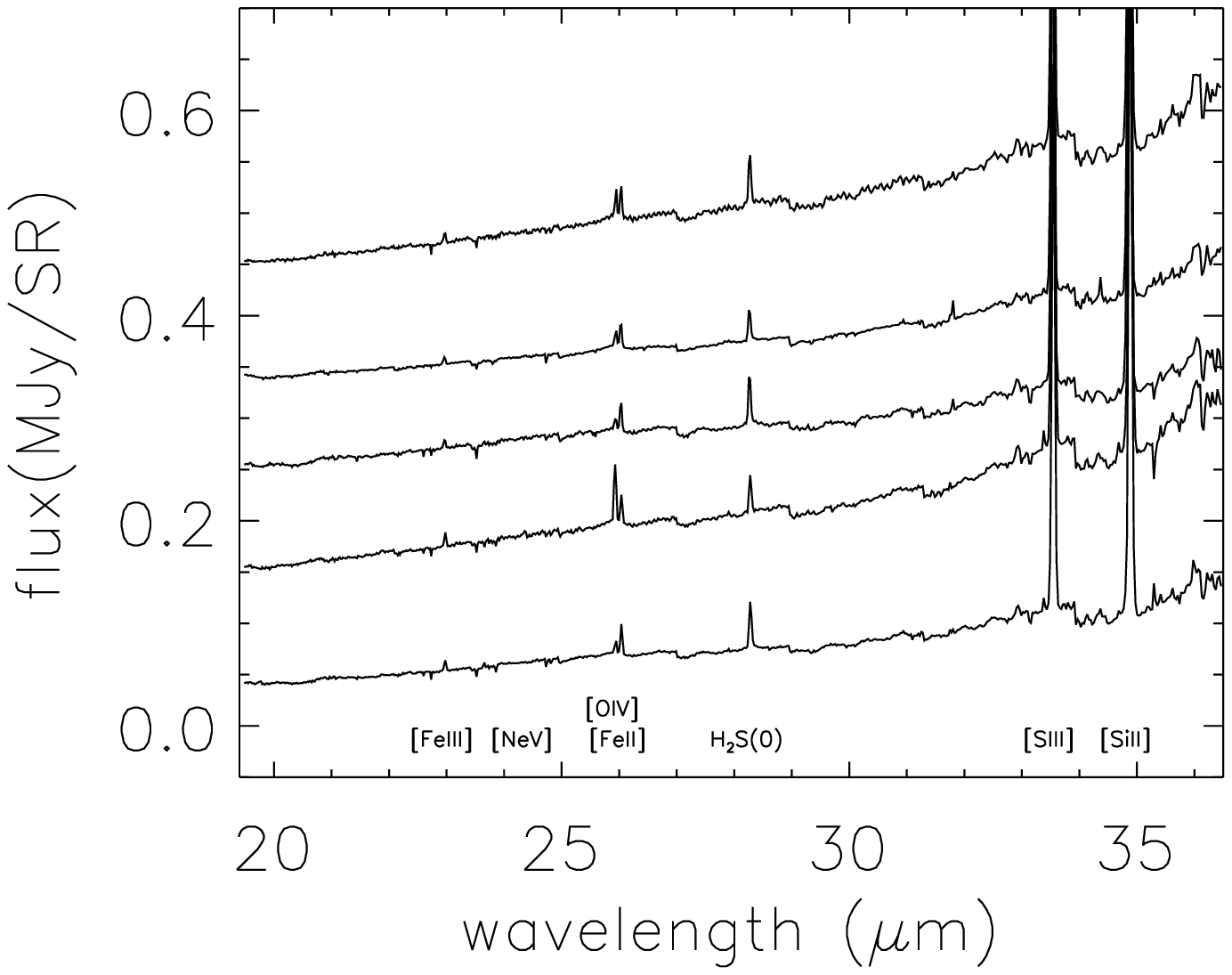}
\includegraphics[angle=0,width=3.5in]{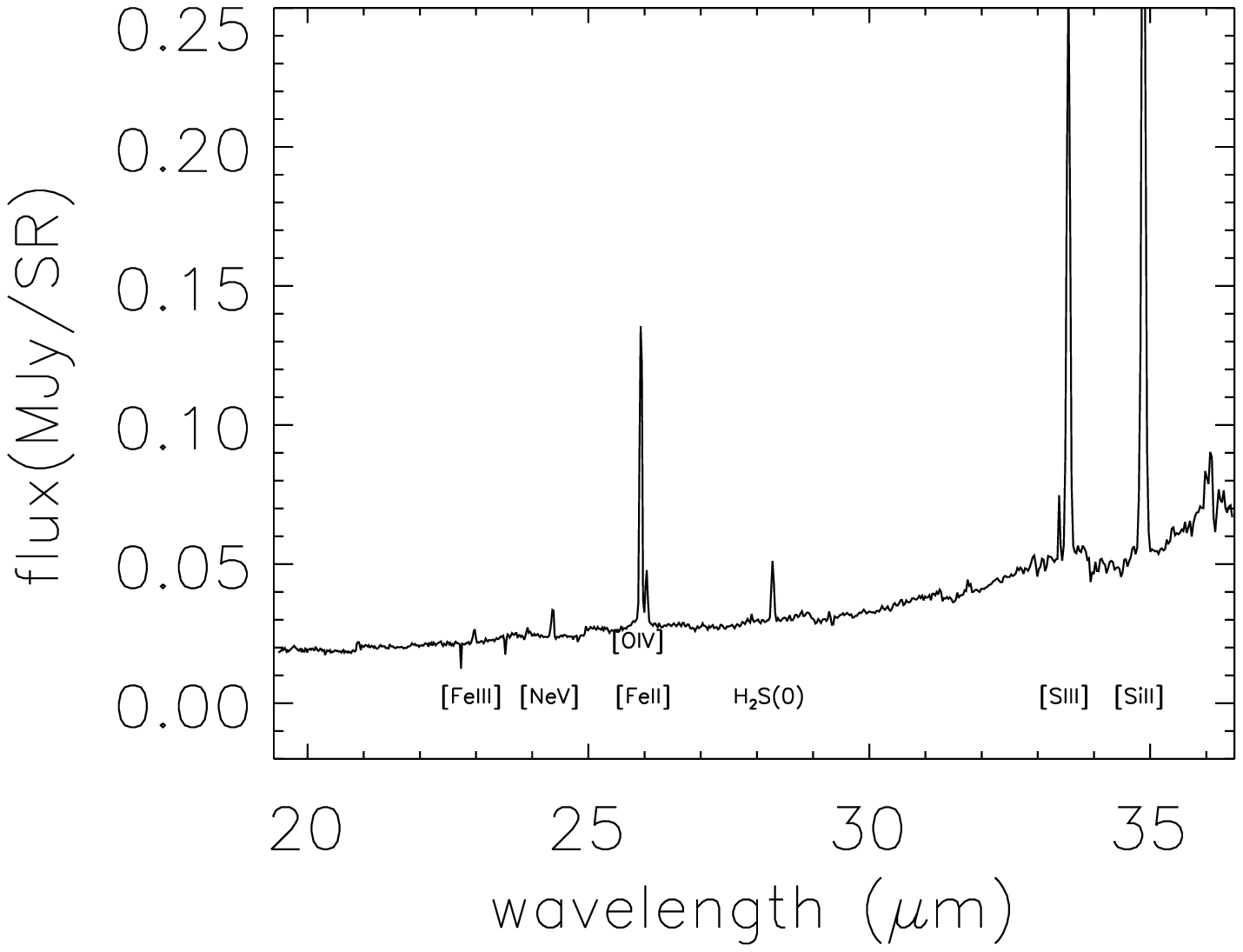}
\caption{
\label{fig:LHspec}
a) 
Spectra from the star forming warped disk 
(seen as a parallelogram in continuum) are shown at the 4 peaks
in the parallelogram and at one additional point
also in the parallelogram.  
Each spectrum is offset by $+$0.1 MJy/SR from the other.  
From bottom to top:
The northern peak west of the nucleus,
the southern peak west of the nucleus,
the southern peak east of the nucleus,
the northern peak east of the nucleus,
a point between the southern and eastern peak and the nucleus.
Positions are given in the text.
In the star forming disk or parallelogram [NeV]($24.318\mu$m) is not detected
and [OIV]($25.890\mu$) is weaker than [FeII]($25.988\mu$m).
b) Spectrum from a region near the jet axis south-west of nucleus.
Near the jet axis [NeV]($24.318\mu$m) is detected and [OIV]($25.890\mu$)
is 3-5 times brighter than [FeII]($25.988\mu$m).
The presence of [NeV] implies suggests that the radiation
field is hard near the jet axis.
}
\end{figure}

In the SL spectral cube we extracted a spectrum
from the same bright region along the jet
axis south-west of the nucleus as that
shown in Figure \ref{fig:LHspec}b and 
from a representative region in the parallelogram
representing the star forming disk.  
The spectrum representing the star forming disk was 
extracted from a position centered at RA=13:25:25.9, DEC=-43:01:09.
The location is the same as that for the second 
from bottom spectrum shown in Figure \ref{fig:LHspec}.
The two spectra extracted from the SL spectral cube 
are shown in Figure \ref{fig:SLspec}.  

As can be seen from Figure \ref{fig:SLspec}a the star forming region exhibits
little molecular hydrogen emission in the higher S rotational lines.
However, the S(2), S(3), and S(5) lines are detected  
near the jet axis and above the 
disk near the dust shell (see Figure \ref{fig:SLspec}b).
The S(7) line is barely detected at   5.511$\mu$m.  
The S(4) line at 8.0241 is not
detected likely because it has been overpowered  by
the dust emission feature at 7.7$\mu$m. 
Likewise the S(6) line at 6.1086$\mu$m is
not detected as it would have been overpowered
by the dust emission feature at 6.2$\mu$m.
The S(5) line is detected above the disk but not in the star forming region
 at 6.909$\mu$m (Figure \ref{fig:SLspec}b).  At almost the same wavelength
the [ArII] line at 6.985$\mu$m is seen in the star forming disk.
The spectra are sufficiently high spectral resolution that we can differ
between the two lines.
A comparison between the two spectra 
also shows that the relative heights of the 
7.7 and 11.3$\mu$m dust emission features
differs.  These two dust emission features are approximately the same
height in the star forming disk.  However  the 11.3$\mu$m emission feature is 
twice as bright as the 7.7$\mu$m dust emission feature in the dust shell.
The [NeII](12.81$\mu$m) and dust emission feature at 11.3$\mu$m are also similar in height
in the star forming region but the 11.3$\mu$m is almost twice as bright
as [NeII] above the disk.   These changes are reflected in the ratio maps
shown in Figure \ref{fig:ratio} and discussed in section 3.

There is a weak feature near 14.2$\mu$m in both spectra
from the star forming disk and jet region (Figure \ref{fig:SLspec}a,b).
The shape of the spectra are similar 
in shape at 14.2$\mu$m suggesting
that this feature is a dust emission feature rather
than due to emission from [NeV](14.3$\mu$m).
A higher spectral resolution spectrum would be required to detect this line.
Weak [NeV](14.3$\mu$m) emission was detected in the 
SH spectrum of the nucleus \citep{weedman05}.

\section{Line ratios }

We have measured line fluxes from the spectra shown
in Figures \ref{fig:LHspec}b and \ref{fig:SLspec}b  which were
extracted at 
the location of the [OIV] peak south-west of the nucleus.
Table \ref{tab:emjet} lists the fluxes of the emission lines
from ions and Table \ref{tab:hjet} lists the fluxes of 
the pure rotational molecular hydrogen transitions.
Line fluxes were measured from the 3 spectral cubes in 
the same aperture using continuum fitting and subtraction
routines available in CUBISM.  We have used CUBISM to measure
line fluxes for the LH spectral cube even though 
the images (displayed in our
figures) show the spectral cube reduced
with the software described by \citet{neufeld06}.
This spectral cube had fewer artifacts than the one produced
by CUBISM, however the artifacts in the CUBISM reduced LH spectral cube should not have
affected the measured line fluxes as
we were able to choose where to measure the continuum and line emission.
Also the same software package was used to measure all
line fluxes to check and ensure consistency in the calibration between the 
spectral cubes.

\subsection{Molecular hydrogen emission}

Pure rotational transitions of molecular hydrogen in the mid-infrared
(between 3 and 20$\mu$m) are a major coolant of
of warm gas in the temperature range 100-1000K \citep{neufeld93}. 
Higher S pure rotation molecular hydrogen lines 
have previously been detected in extragalactic objects.
\citet{higdon06} detected pure rotational 
lines from ULIRGS including the S(7) line in a few objects.  
\citet{panuzzo07} detected these lines  in the elliptical
galaxy NGC 4435 that is experiencing an off-center
encounter with a nearby galaxy. This galaxy hosts a dusty disk as
seen from visible band HST images and ongoing star formation
at modest rate of $\sim 0.1 M_\odot/$yr. 
\citet{roussel07} observed warm gas in the centers of nearby galaxies
with LINERs and Seyferts having warmer H$_2$ emitting in the excited states
or higher S transitions.
The above studies did not resolve the H$_2$ emission or did not study its
morphology.
Higher angular resolution
near infrared studies have found emission from molecular hydrogen
in vibrational transitions in active galaxies. 
In some Seyferts this emission is resolved
on 100pc or larger scales (e.g., \citealt{quillen99,davies05})
in others rotation is seen near the massive black hole 
(e.g., \citealt{hicks07}). 

The flux of a pure rotational transition can be written
$F = {h c \over \lambda} A N_J {\Omega \over 4 \pi}$ 
where $h$ is Planck's constant, $c$ is the speed
of light, $\lambda$ is the wavelength of the transition, $A$ 
is the Einstein A-coefficient, $\Omega$ is the solid angle and $N_J$ 
is the column density of the initial $J$ quantum state.
Since we are consider pure rotational $S$ transitions the rotational 
quantum number change is $\Delta J=-2$,  
the vibrational quantum number change is $\Delta \nu=0$, and $\nu=0$.

\begin{figure}
\includegraphics[angle=0,width=3.5in]{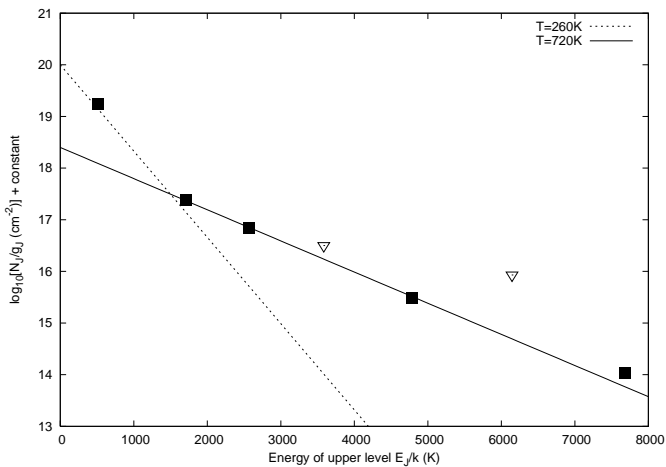}
\caption{
\label{fig:h2plot}
Excitation diagram of the H$_2$ pure rotational lines. 
Points are based on those measured
in the jet region with fluxes listed in Table \ref{tab:hjet}.
Solid squares refer to measurements and open triangles to upper limits.
The solid line is for a temperature of $T_2=720K$ whereas the dotted
one is $T_1 = 260K$.
}
\end{figure}

Assuming local thermal equilibrium, the fraction of molecules
in the $J-$th state is $N_J/N(H_2) = g_J \exp (-E_J/kT)/Z(T)$ where
$Z(T)$ is the partition function.
Here $g_J$ is the statistical weight of the state or  
$2J+1$ times 1 or 3 for the even and odd $J$ states, respectively.
To estimate the temperature of the warm molecular hydrogen we
make an excitation diagram plotting the energy of the initial state $E_J/k$
versus $\log_{10}(N_J/g_J) +$ a constant.  
The excitation diagram is shown in Figure \ref{fig:h2plot}.
On such a plot a single temperature gas with temperature $T$ 
would have points with slope equal to $\log_{10} (e)/T$.
We have chosen to plot using $\log_{10}$ rather than the natural
log  so 
that the $y$-axis gives an estimate for the column density of the states.

Figure \ref{fig:h2plot} shows that a single temperature model
does not fit all data points.  This is not unexpected as other
extragalactic objects also require multi-temperature models 
(e.g., \citealt{higdon06,panuzzo07,roussel07,ogle07}).
The points lie along a line for initial states higher than $J=2$
implying that the higher $J$ rotational level populations are
close to LTE.
The S(2) and higher lines are fit with a temperature of $T_2=720K$.
The S(0) and S(2) line are connected with a slope corresponding
to a temperature of $T_1 \sim 260K$. 
Had we subtracted the warmer component, a somewhat
lower temperature would have been measured.  Unfortunately our
spectra do not cover the S(1) line. This would have helped in
making a 2 component model.
It is unlikely that there is a strong
deviation from thermalization
of ortho levels with para levels as the S(2) point lies along
the same line as the S(3), S(5) and S(7) points in the
excitation diagram.  

The column density corresponding to the above excitation model
is $N(H_2)\sim 10^{20}$cm$^{-2}$ for the cooler molecular component ($T_1=260K$)
and $\sim 10^{19}$cm$^{-2}$ for the warm component ($T_2=720K$). The column densities
are similar to those measured in the nuclei of nearby galaxies by \citet{roussel07} and in supernova remnants where blast waves encounter molecular clouds 
\citep{neufeld07}.
The temperatures of the two components we measure are higher than most of those measured 
by \citet{roussel07} for
most nearby galaxy centers excepting the LINERS and Seyfert galaxies,
but lower than the supernova remnants \citep{neufeld07} and
ULIRGs with two temperature models \citep{higdon06}.  
The fraction, $\sim 10^{-1}$, in the warmer component (that at 720K compared
to that at 260K) is higher than that
measured for the ULIRGS  with 2 component temperature models and most
but not all the galaxies studied by \citet{roussel07} but
similar to that measured from the supernova remnants \citep{neufeld07}.
The physical conditions in Cen A's dustshell are most closely matched
by the supernovas studied by \citet{neufeld07}.

As we have discussed from the images, the molecular hydrogen
emission in the higher S lines arises from the dust shell and along
the jet axis, though
that in S(0) is associated with the disk
and so is probably from with photodissasociation regions (PDRs) associated
with star formation.
We did not clearly detect the higher S lines in the disk suggesting
that the molecular hydrogen at the hotter $T_2 = 720K$ temperature
is primarily excited in the vicinity of the dust shell or along the
jet axis.
The column density of cooler molecular hydrogen is similar
to that estimated in the dustshell by \citet{quillen_supershell} 
from the shell surface brightness.  
This suggests that gas
associated with the dust shell has been heated, in the shell and
more so near the jet axis.
There is probably more than one excitation process as [OIV] and [NeV]
emission are detected primarily along the jet axis whereas
the higher S pure rotational molecular hydrogen lines are detected
there and in the vicinity of the dust shell.

\subsection{Excitation near the jet axis}

Nebular emission lines can constrain the 
the spectrum of the energy powering the emission.
Our spectra allow us to make use of the following line
ratios: [OIV]$25.9$/[SIII]$33.5\mu$m, and [OIV]25.9/[NeII]12.8$\mu$m.
We do not detect [NeV] at $14.3\mu$m so we will discuss limits
on the ratios [NeV]14.3/[NeII]12.8$\mu$m and [NeV]14.3/[NeV]24.3$\mu$m.

We measure  [OIV]$25.9$/[SIII]$33.5\mu$m $\sim 1.0$ at the [OIV] peak.
This ratio is most similar to those of the AGNs considered by
\citep{genzel98} (see their figure 3).
However we have failed to detect a significant [NeV] line at 14.3$\mu$m so
we estimate a limit on [NeV](14.3)/[NeII](12.8) $< 0.05$. 
Ratios this low are more similar to starbursts than AGNs.
We note that the ratio we measure in this region 
[OIV]25.9/[NeII]12.8 $\sim 0.17$ is lower than expected
from the [OIV]25.9/[SIII]33.5 ratio using 
Genzel et al.'s corrective factor of 1.7.
Our estimated [OIV]25.9/[NeII]12.8 $\sim 0.2$ ratio places the emission along
the jet axis more similar to ULIRGs than AGNs.
[OIV]25.9/[NeII]12.8 is expected to be approximately 1 for AGNs 
(\citealt{genzel98}, Figure 3a
and Figure 7 by \citealt{sturm02}).
Our low estimated ratios of [NeV]14.3/[NeII]12.8$\mu$m 
and  [OIV]25.9/[NeII]12.8$\mu$m
compared to those expected in AGNs may in
part be due to the blending of the [NeII]12.8 emission 
with dust emission features which may have caused us to overestimate 
the [NeII](12.8$\mu$m) line strength.

We don't detect [NeV](14.3$\mu$m) or [NeVI] at 7.6$\mu$m and we do
detect [NeV](24.3$\mu$m).   By comparing upper limits of ratios of these lines 
with [OIV](25.9) and [NeV](24.3$\mu$m) and
Figure 4 by \citet{sturm02} (based on models by \citealt{spinoglio00}) we infer
the emitting region is likely to have low ionization parameter
$U \lesssim 10^{-2}$. 
Our estimated upper limit
in the ratio of [NeV]14.3/24.3 of $<3$ in the jet region and a comparison
of Figure 3 by \citet{alexander99} suggests that the electron density
is also low $n_e \lesssim 10^2$cm$^{-3}$.
The detection of [NeV] suggests that a hard radiation field is illuminating
the dust shell near the jet axis. However we have failed  to detect
[NeV](14.3$\mu$m) or [NeVI](7.65$\mu$m) suggesting that the illuminated region
is low density and has a low ionization parameter.
The difference in the physical conditions compared to those expected in
narrow line regions might account for the discrepancy between
the [OIV]25.9/[NeII]12.8  and [OIV]25.9/[SIII]33.5 ratios compared
to those seen in AGNs.

We consider the possibility that 
the low electron density and ionization parameter are consistent
with illumination of a region distant from the nucleus
by hard radiation from the central AGN.
The position of the [OIV](25.9$\mu$m) peak corresponds to a distance 
$d \sim 400$ pc from the nucleus.
The active nucleus is estimated to have a bolometric luminosity of
$L_{bol} \sim 10^{43}$ erg/s \citep{whysong04}.
The ionization parameter, $U$, for isotropic emission is
$U \equiv {Q \over 4 \pi d^2 n c}$ where $c$ is the
speed of light, $Q$ is the ionizing photon emission rate (photons per second)
and $n$ is the gas density (in cm$^{-3}$).
Assuming a mean energy for ionizing photons of
100ev (the estimated energy of the UV bump and hard enough
to produce NeV from the preceding ionization state) 
at the estimated bolometric luminosity 
emitted isotropically into $4\pi$ radians, 
$$ U \sim 0.001 {\rm cm}^{-3}
\left({L_{bol}\over 10^{43}{\rm erg~s^{-1}}}\right)
\left({d \over 400 {\rm pc}}\right)^{-2}
\left({n \over 10^{2} {\rm cm}^{-3}}\right)^{-1}.
$$
This estimate illustrates that a low ionization parameter, low density
medium (such as we infer from the line ratios) is the only
situation that we might expect at a distance of 400~pc from the AGN.

We now ask: can the AGN nucleus 
provide enough UV photons to account for the flux of the [OIV] emission 
25\arcsec from the nucleus?
We summed the [OIV](25.9$\mu$m) flux in a 15\arcsec wide region centered 
on the [OIV] emission
peak, estimating a flux of $10^{-13}$~erg~cm$^{-2}$~s$^{-1}$ in the line. 
This flux corresponds to a luminosity in [OIV] of $10^{38}$ erg/s.  
We can compare the [OIV] luminosity to the mid-infrared luminosity
as have \citet{sturm02} for other objects.
Since the [OIV] flux is illuminated only in a small
solid angle from the nucleus we must first estimate
the fraction of UV luminosity that could
be causing the [OIV] emission.
The [OIV] emission comes from a region that is about 10\arcsec wide. 
As viewed from the position of the nucleus, this corresponds to a
solid angle of $(20^\circ)^2$ or $\sim 1$\% of $4\pi$ steradians.
Thus we compare the [OIV] luminosity to 1\% of the mid-infrared luminosity.
The mid-infrared luminosity of
the AGN is $\sim 10^{42}$erg~s$^{-1}$ \citep{whysong04}.
Consequently we estimate that 
the ratio of the [OIV] to 1\% of the mid-Infrared luminosity is $\sim 10^{-2}$.
This ratio is similar to that seen for the brighter Seyferts studied by 
\citet{sturm02} (see their Figure 12).
We conclude that as long as they are not absorbed before
they run into the dust shell, UV photons from the AGN 
can account for the [OIV] and [NeV] emission near the jet axis.
An additional power source (like shocks associated with the jet)
is not necessarily required to account for the high ionization species
seen near the jet axis but distant from the nucleus.
If UV photons from the AGN are responsible for the high
ionization species then the column depth along the sight line between
the AGN and dust shell must be low.
If a blast wave is responsible for sweeping up the dust shell, 
as proposed by \citet{quillen_supershell}, then it could also 
be responsible for evacuating material between
the AGN and the dust shell.


\section{Summary and Conclusion}

We have carried out a spectroscopic study of the central 2 arcminutes of
Centaurus A using short low and long high spectral modules of the Infrared
Spectrograph on board the {\it Spitzer Space Telescope}.  
Most of the emission lines detected in the spectral cubes
(e.g., [SIII](33.5$\mu$m), [SiII](34.8), [FeII](26.0), [FeIII](23.9), [ArII](6.98) and H$_2$S(0)(28.2)) and dust emission features
primarily trace regions of star formation in the warped disk.

Our previous study based on IRAC and MIPS imaging
suggested that Centaurus A hosts an oval or bipolar dustshell at
a distance approximately 500 pc  from the nucleus seen
above and below the warped disk.
This dust shell, if confirmed, would be
the first extragalactic shell to be discovered in the infrared.
Here we see the dust shell even more clearly and prominently
in the 11.3$\mu$m dust emission feature than we saw
previously in the broad band IRAC images.
We have found variations in the dust emission 
feature 7.7$\mu$m/11.3$\mu$m ratio
and dust 11.3$\mu$m/[NeII](12.8$\mu$) ratio,
with the oval dust shell having the lowest
ratios compared to the star forming disk.
The clearer shell morphology at 11.3$\mu$m than previously
seen in broad band images,
the association of the molecular hydrogen emission in the shell,
and the variation in line ratios in the shell compared to
those in the disk, confirm spectroscopically that 
the shell discovered previously \citep{quillen_supershell} 
is a separate coherent entity and is unlikely to
be a chance superposition of dust filaments.

We find evidence for higher ionization species line emission in 
[NeV]($24.3\mu$m) and [OIV]($25.9\mu$m) near the jet axis.  
Emission in these two lines is seen
both north-east and south-west of the nucleus along position angles 
$\sim 40^\circ$ and $\sim -120^\circ$.  These angles are similar to but not
exactly the same as the jet axis at 55$^\circ$ as seen at 5GHz. 
Outside the nucleus, the peak surface brightness
in these lines is 25\arcsec or 400~pc south-west of the nucleus.  
Emission line ratios and limits at the location of the [OIV] peak
suggest that the emitting region is at low ionization parameter, 
$U \lesssim 10^{-2}$,  and has low electron density,
$n_e \lesssim 10^2$cm$^{-3}$.
We crudely estimate that the AGN can provide sufficient UV photons 
to account for the [OIV] luminosity 400~pc from the nucleus,
as long as UV photons are not absorbed by intervening
material as they travel from the AGN to the dust shell.
A much more detailed photo-ionization study is required to
understand the excitation of the [OIV] and [NeV] emission.
Previous reports of an ionization cone
in Cen A were based on near-infrared imaging of
the central few arcseconds \citep{bryant99}.  
Subsequent studies interpreted line emission in 
terms of a disk rather than ionization cone 
\citep{schreier98,krajnovic06}.
Unfortunately
the morphology of the near-infrared images is strongly affected by extinction
and the warp disk models have not been good enough to accurately
predict the extinction in the central few arcsecond of the nucleus. 
Cen A might be the only active galaxy 
in which mid-infrared spectroscopy
has found evidence for high ionization species such
as NeV at hundred pc distances from the nucleus.
As far as we know Cen A hosts the only ionization cone 
that has been resolved with observations from the Spitzer
Space Telescope.


We see evidence for warm molecular hydrogen coincident 
with the peak in [OIV]
in the odd pure rotational odd transitions S(3) and S(5).
The S(7) and S(2) transitions are also detected but at weaker levels.
The S(3) and S(5) emission also lies in the vicinity of the dust shell that is
most prominent in the 11.3$\mu$m dust (PAH) emission feature.
A two temperature component model can fit the rotational line ratios and implies
that there is warm molecular hydrogen with temperatures in the range 250--720K.
The temperatures are warmer than seen in nucleus of non-active nearby galaxies,
similar to those of LINERS and Seyferts but lower than exhibited by 
supernova remnants ULIRGS (as
compared to studies by \citealt{roussel07}, \citealt{higdon06} and
\citealt{neufeld07}).
Near the jet axis, the column depth of warm molecular hydrogen 
is $N(H_2) \sim 10^{20}$cm$^{-2}$
similar to that estimated from the infrared continuum emission of the dust shell
by \citet{quillen_supershell}.
This suggests that gas associated with the dust shell has been heated 
near the jet axis.
There is probably more than one excitation process as [OIV] and [NeV]
emission are detected primarily along the jet axis whereas
the higher S pure rotational molecular hydrogen lines are detected
there and in the vicinity of the dust shell.

Previous studies of the pure-rotational molecular hydrogen lines
in extra-galactic objects (e.g., \citealt{higdon06,panuzzo07,roussel07,ogle07}) 
have not well resolved the emission.
The association of the warm molecular hydrogen gas with a shell
is most similar to phenomena exhibited by Galactic supernova remnants 
where the blast wave encounters molecular clouds \citep{neufeld07}.
The physical conditions estimated from the molecular hydrogen observations
are similar in properties (column depth, temperatures and fraction of 
gas in the two temperature components) to the
parameters estimated by \citet{neufeld07} in 4 Galactic
supernova remnants.
This suggests that theory of interstellar shock waves be applied
to interpreting observations of the shell.  
\citet{neufeld07} associates
the warmer molecular hydrogen component responsible for 
the higher S pure rotational transitions with dissociative shocks.
These require shock velocities 
$\gtrsim 70$km/s (e.g., \citealt{drainemckee93}).
A physical scenario and model 
accounting for the shell's structure and energetics is currently lacking.   
Deep optical spectroscopic and radio studies are particularly needed
to better constrain gas motions and physical conditions in this shell. 


\vskip 1.0truein
We thank Dan Watson, Paul van der Werf,  Ralph Kraft, Jacqueline van
Gorkom, Martin Hardcastle,  and Christine Jones-Forman
for helpful discussions and correspondence.
We thank Martin Hardcastle for providing us with images
of Centaurus A in the radio.
Support for this work was in part provided by 
by NASA through an award issued by JPL/Caltech,
National Science
Foundation grants AST-0406823 $\&$ PHY-0552695, the National
Aeronautics and Space Administration under Grant No.$\sim$NNG04GM12G
issued through the Origins of Solar Systems Program, and
HST-AR-10972 to the Space Telescope Science Institute.
JBH is funded by a Federation Fellowship from the Australian Research
Council.

\clearpage

\clearpage

\clearpage

\clearpage

\clearpage

{}

\begin{table*}
\begin{minipage}{120mm}
\caption{Lines Detected \label{tab:lines}}
\begin{tabular}{@{}ll}
\hline
Line  & Rest Wavelength($\mu$m)  \\ 
\hline
[FeII]         & 5.340   \\
H$_2$S(7)J=9-7 & 5.511   \\
Dust           & 5.7     \\
Dust           & 6.2     \\
H$_2$S(5)J=7-5 & 6.909   \\
$[$ArII$]$     & 6.985   \\
Dust           & 7.7     \\
Dust           & 8.6     \\
H$_2$S(3)J=5-3 & 9.665   \\
$[$SIV$]$      & 10.51   \\
Dust           & 11.3    \\
Dust           & 12.0    \\
H$_2$S(2)J=4-2 & 12.279  \\
Dust           & 12.7    \\
$[$NeII$]$     & 12.81   \\
Dust           & 14.2    \\
\hline
$[$FeIII$]$    & 22.925  \\
$[$NeV$]$      & 24.318  \\
$[$OIV$]$      & 25.890  \\
$[$FeII$]$     & 25.988  \\
H$_2$S(0)J=2-0 & 28.219  \\
$[$SIII$]$     & 33.481  \\
$[$SiII$]$     & 34.815  \\
\hline
\end{tabular}
{ \\
The spectra in the star forming disk and jet region have the same
shape near 14.2$\mu$m suggesting that the feature at
that wavelength is a dust emission feature \citep{smith07}
rather than due to emission from [NeV](14.3$\mu$m).
The H$_2$S(4)J=6-4 line at 8.025$\mu$m is not detected. The emission
at this wavelength is dominated by 
nearby bright 7.7 and 8.6$\mu$m dust emission features. 
The S(6) line at 6.1086$\mu$m is also
not detected as it would have been overpowered
by the dust emission feature at 6.2$\mu$m.
The even $J$ molecular hydrogen rotation quantum states are singlets
rather than triplets and so should be similar to 3 times fainter than
nearby odd $J$ transitions.
At the [OIV] and H$_2$S(3) peak south-west of the nucleus the
[ArII](6.985$\mu$m) is weaker than the H$_2$S(5)(6.909$\mu$m) line.  
The opposite is true in the star forming disk.
[FeII](5.3$\mu$m is detected in the disk and
at the OIV peak near the jet axis.
The line in the middle above list denotes the break in wavelength
between the SL and LH wavelength coverage.
}
\end{minipage}
\end{table*}

\begin{table*}
\begin{minipage}{120mm}
\caption{Nebular Line Fluxes at [OIV] peak\label{tab:emjet}}
\begin{tabular}{@{}ll}
\hline
Line & Flux  \\
\hline
$[$OIV$]$(25.9$\mu$m)  &  1.5 \\
$[$NeII$]$(12.8$\mu$m) &  9.0 \\
$[$NeV$]$(24.3$\mu$m)  &  0.15 \\
$[$NeV$]$(14.3$\mu$m)  & $<$ 0.5 \\
$[$SIII$]$(33.5$\mu$m) &  1.6 \\
$[$SiII$]$(34.8$\mu$m) &  4.0 \\
$[$SIV$]$(10.5$\mu$m) &  1.6 \\
$[$FeII$]$(26.0$\mu$m)  &  0.2   \\
$[$FeIII$]$(22.9$\mu$m) &  0.05 \\
$[$ArII$]$(6.98$\mu$m)  &  $\sim$0.1   \\
\hline
\end{tabular}
{ \\
The fluxes were measured in a square region
centered at RA=13:25:25.7 DEC=-43:01:24 and 11\arcsec wide.
that is a peak in [OIV] and H$_2$S(3) emission (see figure \ref{fig:h2s3_ov}).
The spectra from this region are shown in Figures 
\ref{fig:LHspec}b and \ref{fig:SLspec}b.
Fluxes are given in units of $10^{-5}{\rm erg~cm}^{-2}{\rm s}^{-1}{\rm SR}^{-1}$.
The [NeV](14.3$\mu$m) flux is an upper limit.
We did not detect [NeVI] at 7.642$\mu$m.
}
\end{minipage}
\end{table*}

\begin{table*}
\begin{minipage}{120mm}
\caption{Pure rotational H$_2$ Line Fluxes at [OIV] peak\label{tab:hjet}}
\begin{tabular}{@{}ll}
\hline
Line & Flux \\ 
\hline 
S(0)(28.2$\mu$m) &  0.21 \\
S(2)(12.2$\mu$m) &  1.1 \\
 S(3)(9.7$\mu$m) &  5.2 \\
 S(4)(8.0$\mu$m) & $<$3 \\
 S(5)(6.9$\mu$m) &  2.7 \\
 S(6)(6.1$\mu$m) & $<$6 \\
 S(7)(5.5$\mu$m) &  0.5  \\
\hline
\end{tabular}
{\\
Fluxes were measured in the same region as listed in Table \ref{tab:emjet}
and in the spectra shown in Figures \ref{fig:LHspec}b and \ref{fig:SLspec}b.
Fluxes are given in units of $10^{-5}{\rm erg~cm}^{-2}{\rm s}^{-1}{\rm SR}^{-1}$.
For the S(4) and S(6) lines we estimate upper limits.
The upper limits are high because of dust emission features.
}
\end{minipage}
\end{table*}

\end{document}